\documentclass[reprint,showpacs,preprintnumbers,amsmath,amssymb,a4paper,nofootinbib,superscriptaddress,floatfix,aps,rmp,showkeys]{revtex4-1}

\setcitestyle{authoryear, round, semicolon, open={(}, close={)}, aysep={, }}

\DeclareRobustCommand{\baselinestretch{1.0}}

\usepackage{lineno}
\usepackage{blindtext, color}

\usepackage{graphicx}

\usepackage{dcolumn}
\usepackage{bm}
\usepackage{amsmath}
\usepackage{todonotes}
\usepackage{placeins} 

\usepackage{color}
\usepackage{colortbl}

\usepackage{txfonts}
\usepackage[utf8]{inputenc}

\usepackage{xcolor}
\usepackage[colorlinks=true, pdfstartview=FitV, linkcolor=blue, citecolor=blue, urlcolor=blue]{hyperref}

\makeatother

\begin{document}

\preprint{$\today$}

\title{Fast Pixelated Detectors in Scanning Transmission Electron Microscopy. Part II: Post Acquisition Data Processing, Visualisation, and Structural Characterisation}

\author{Gary W. Paterson}
  \email{Dr.Gary.Paterson@gmail.com}
  \affiliation{SUPA, School of Physics and Astronomy, University of Glasgow, Glasgow G12 8QQ, UK.}
\author{Robert W. H. Webster}
  \affiliation{SUPA, School of Physics and Astronomy, University of Glasgow, Glasgow G12 8QQ, UK.}
\author{Andrew Ross}
  \altaffiliation[Current address: ]{Institute of Physics, Johannes Gutenberg Universit\"{a}t Mainz, Staudingerweg 7, 55129 Mainz, Germany.}
  \affiliation{SUPA, School of Physics and Astronomy, University of Glasgow, Glasgow G12 8QQ, UK.}
\author{Kirsty A. Paton}
  \affiliation{SUPA, School of Physics and Astronomy, University of Glasgow, Glasgow G12 8QQ, UK.}
\author{Thomas A. Macgregor}
  \affiliation{SUPA, School of Physics and Astronomy, University of Glasgow, Glasgow G12 8QQ, UK.}
\author{Damien McGrouther}
  \affiliation{SUPA, School of Physics and Astronomy, University of Glasgow, Glasgow G12 8QQ, UK.}
\author{Ian MacLaren}
  \affiliation{SUPA, School of Physics and Astronomy, University of Glasgow, Glasgow G12 8QQ, UK.}
\author{Magnus Nord}
\email{Magnus.Nord@ntnu.no}
  \altaffiliation{Current address: Department of Physics, Norwegian University of Science and Technology, Trondheim 7491, Norway.}
  \affiliation{SUPA, School of Physics and Astronomy, University of Glasgow, Glasgow G12 8QQ, UK.}
  \affiliation{EMAT, Department of Physics, University of Antwerp, Antwerp 2000, Belgium.}

\date{\today}

\begin{abstract}
Fast pixelated detectors incorporating direct electron detection (DED) technology are increasingly being regarded as universal detectors for scanning transmission electron microscopy (STEM), capable of imaging under multiple modes of operation.
However, several issues remain around the post acquisition processing and visualisation of the often very large multidimensional STEM datasets produced by them.
We discuss these issues and present open source software libraries to enable efficient processing and visualisation of such datasets.
Throughout, we provide examples of the analysis methodologies presented, utilising data from a 256$\times$256 pixel Medipix3 hybrid DED detector, with a particular focus on the STEM characterisation of the structural properties of materials.
These include the techniques of virtual detector imaging; higher order Laue zone analysis; nanobeam electron diffraction; and scanning precession electron diffraction.
In the latter, we demonstrate nanoscale lattice parameter mapping with a fractional precision $\le 6\times10^{-4}$ (0.06\%).
\end{abstract}

\keywords{fast pixelated detector, 4D-STEM, virtual detectors, higher order Laue zone, scanning precession electron diffraction}

\maketitle

\section{Introduction}

One of the greatest revolutions in scanning transmission electron microscopy (STEM) in recent years is the development and use of fast pixelated detectors incorporating direct electron detection (DED) technology, and these are rapidly becoming a key component of the imaging system for a modern STEM.
\citealp{ophus_mm_2019_4dstem} has provided an excellent review of the area, and \citealp{tate_2016_mandm_empad}, \citealp{Yang_2015_phase_contrast}, and \citealp{matus_pixelated_stem_magnetic_2016} have described some suitable detectors for different applications in pixelated STEM imaging.
Part~I \cite{fpd_part1_arxiv} of this work briefly discussed the general benefits of fast pixelated detectors for use in STEM, and presented software solutions for their hardware control, and for the acquisition, real-time processing and visualisation, and storage of data from them.
An increasing number of Python packages are being developed by the electron microscopy community with capability to process four-dimensional (4-D) STEM (4D-STEM) data, including HyperSpy \cite{hyperspy_library}, LiberTEM \cite{libertem}, py4DSTEM \cite{py4dstem, Savitzky_2020_py4dstem_arxiv}, pycroscopy \cite{pycroscopy}, and pyXem \cite{pyxem}.

In this present article, Part~II, we discuss post-acquisition data exploration and analysis of a variety of STEM datasets acquired with a fast pixelated detector, using the \texttt{fpd} \cite{fpd, fpd_demos} and \texttt{pixStem}\footnote{Towards the very end of preparing this paper, it was decided to merge \texttt{pixStem} with pyXem \cite{pyxem}. All of the features detailed in part~I and II of this series of papers that are related to \texttt{pixStem} are in the processes of being added to pyXem and will continue to be available. The features of the \texttt{fpd} library remain unaffected.} \cite{pixstem} Python libraries, with a focus on their use in mapping the structural properties of materials.
These include the techniques of virtual detector imaging \cite{Rauch_2005_virt_det, Gammer_UM_2015_virt_ap}, higher order Laue zone STEM (HOLZ-STEM) \cite{Huang_holz_stem_2010,Nord_2018_holz}, nano-beam electron diffraction (NBED) \cite{nbed_beche_2009}, and scanning precession electron diffraction (SPED) \cite{Vincent_1994_UM_sped}.
Throughout, we provide examples of the application of these techniques to data recorded with a Medipix3RX hybrid counting DED detector (henceforth referred to as Medipix3) \cite{Ballabriga_2013_Medipix3RX} mounted in STEMs with 200~kV electron sources, though the techniques and software packages discussed here extend to data from other detectors.
The names of the specific functions, classes, methods, modules, and packages used in the examples given are specified in the main text or the figure captions in \texttt{typewriter} font.

Both of the software libraries presented are made available under the free and open source GPLv3 license, allowing transparency of the implemented algorithms, and the ability for anyone to use and to further improve upon them.
The libraries themselves draw on the rich ecosystem of mature Python libraries \cite{oliphant_07_python}, including ones for optimised numerical \cite{numpy} and scientific \cite{scipy} computing, image processing \cite{scikit_image, Gouillart_2016_python_scikit}, and data visualisation \cite{matplotlib}.

The \texttt{pixStem} library is built upon HyperSpy and extends its capabilities to common pixelated STEM tasks, with a focus on processing higher order Laue zone (HOLZ) data.
A key difference between the \texttt{pixStem} and \texttt{fpd} packages is that the latter library is a lower level one, similar in style to NumPy but with added classes, GUIs and plotting capabilities built-in where needed.
A higher level object oriented interface will be added in a future version.
A detailed description of every feature of the \verb|pixStem| and \verb|fpd| libraries can be found in the documentation and example Jupyter \cite{Kluyver_2016_jupyter} notebooks available online \cite{pixstem, fpd, fpd_demos} and will not be replicated here.
Instead, in the following sections, we describe some of the general techniques employed by the libraries for efficient data processing, for data visualisation, and the main features of the libraries for the structural characterisation of materials.

This paper is organised as follows. 
Section~\ref{sec:general_techniques} outlines general techniques for handling post-processing of the large 4D-STEM datasets typically produced by fast pixelated detectors.
In Section~\ref{sec:data_visualisation}, ways of visualising the datasets are discussed.
Section~\ref{sec:virtual_detectors} covers various methods of generating virtual images from the 4-D datasets, using annular dark-field (ADF) contrast from a non-crystalline soft material as an example.
In Section~\ref{sec:holz}, HOLZ-STEM data analysis is explained, which allows for the retrieval of information about the periodicity of a crystalline sample in the direction parallel to the electron beam.
Lastly, general lattice image processing is discussed in Section~\ref{sec:lattice}, with a focus on diffraction imaging.
Using test data obtained in the scanning precession electron diffraction (SPED) mode of acquisition \cite{Vincent_1994_UM_sped} of a custom system \cite{MacLaren_MandM_2020_sped}, we demonstrate that a lattice parameter fractional precision of $6\times10^{-4}$ is possible using a probe with a spatial resolution of 1.1~nm.
This precision value is approximately a factor of 2-3$\times$ higher than the best ones reported in the literature using standard probes in SPED mode \cite{Rouviere_APL_2013_SPED} and patterned probes in standard STEM mode \cite{Guzzinati_2019_bessel_arxiv, Zeltmann_UM_2020_patterned_aps}.

A future part~III of this work will cover post-acquisition processing and visualisation of data from fast pixelated detectors for differential phase contrast imaging.

\section{General Techniques}
\label{sec:general_techniques}
The primary challenge to the post acquisition processing of data from fast pixelated detectors is the size of the data, which can easily be much greater than the available system memory on current generation computers.
For example, a spatial scan with 1k$\times$1k points recording data from a 512$\times$512 pixel detector would occupy $\sim$977~GB when stored in 32-bit integer format.
In addition to this, in some types of processing, at least the same amount of memory is required to process the data, putting the potential memory requirements well into the terabyte range.
For STEM imaging, smaller scan sizes of 256$\times$256 or 512$\times$512 points with a 256$\times$256 pixel detector at lower bit depth are often adequate to observe the specific feature of interest.
In these cases, and for 16-bit data, the memory requirements are more modest, being 8.6~GB and 34.4~GB (in SI units), but the problem of available system memory and efficient computation currently remains.

One solution to these issues is out-of-core processing, where the data is stored on disk and only parts of it are loaded into memory when needed.
As an example, to generate a bright-field image from a STEM dataset, each reciprocal space image may be loaded one at a time from a hard drive into memory, a sum of each image performed, with the resulting single values stored in memory, and the diffraction pattern memory reused to store the next image.
This is an extreme example and, in reality, multiple images may be loaded into memory at once and processed across multiple CPU cores in parallel to achieve significant performance improvements.

Out-of-core processing is termed `lazy-loading' in HyperSpy and was recently added through the use of Dask \cite{dask_library}, a library which abstracts away the complexities of out-of-core processing.
The \texttt{pixStem} library relies on HyperSpy for out-of-core processing, whereas the \texttt{fpd} library was implemented before out-of-core processing was available in HyperSpy and so implements its own methodology, relying on HyperSpy for Gatan Digital Micrograph file access.

The two libraries can both also process in-memory data.
When processing out-of-core data on disk, there are several ways of making the data exploration and analysis more efficient.
Chunking the dataset in a compressed HDF5 \cite{hdf5_file_format} file, as discussed in Part~I \cite{fpd_part1_arxiv}, in the scan and detector dimensions by partitioning the diffraction pattern data into several pieces, can greatly speed up data processing by both reducing the data reading and decompression time, and by reducing the in-memory array sizes.
For example, if one wants to make a bright-field image by using a small virtual aperture, most of the diffraction pattern can be ignored.
Without chunking the detector dimensions, each full image and, thus, the entire dataset must be read.
However, when a 256$\times$256 pixel diffraction pattern is partitioned into 16$\times$16 chunks and the bright-field disc is located within one such chunk, only 0.39\% of the dataset has to be read and processed.
The use of HDF5 files for data storage brings with it many additional benefits, and these are discussed more fully in Part~I \cite{fpd_part1_arxiv}.

In both packages, reading of chunked data from disk may be aligned to the chunk size when possible to optimise read times.
The size of the data read into memory can be set to be independent of the chunk size or be multiples thereof, allowing data to be processed even on systems with limited memory and processing power.
Additionally, many of the algorithms in the \texttt{fpd} and \texttt{pixStem} library can be set to operate using one CPU core or, by default, to employ all available cores to process the in-memory data chunk in parallel, enabling significant speed increases over single core processing.

Binning the data is another implemented method that can increase data processing speed.
Detector resolution is an important factor when considering the influence of binning on data analysis. 
Most DEDs have relatively few pixels compared to conventional CCD based detectors and the reduction in pixel counts on rebinning can limit the ability to extract the maximum signal from the data.
In many cases, however, modest rebinning can be applied without significant changes in the signal-to-noise ratio (SNR), or large rebinning can be applied to vastly increase the processing speed.
If the data were recorded using a detector that has no readout noise and that counts electrons only in the pixel in which they entered the detector's sensor, then the noise would be Poissonian and, for some analyses, such as centre of mass, rebinning has almost no affect on the SNR of the analysis result.

The Medipix3 detector \cite{Ballabriga_2013_Medipix3RX} has a noise-free readout when the counting threshold is set above the level of the detector's intrinsic noise and, at electron beam energies up to 80~keV, the detector is capable of a near perfect response, counting electrons mainly in the pixel of entry \cite{MIR_2017_medipix3_characterisation}.
However, 200~keV electrons can travel long distances in the sensor and are counted by multiple pixels \cite{McMullan2007_mpx2_electron_imaging}, giving rise to a point spread function of a few pixels.
This correlation modifies the noise profile of the detector from pure Poisson statistics.
However, rebinning the data, even by a small extent, nullifies the correlated component of the noise, resulting in a Poissonian noise profile at the expense of lower detector resolution.
While rebinning tends to have little effect on the SNR of the result of many analysis techniques, it can improve it in some cases, and the speed improvements it brings are particularly useful when performing live data-processing, as discussed in Part~I \cite{fpd_part1_arxiv}, or for rapid post-acquisition assessment of the data while on the microscope.

The binning of data can also be used to reduce the dataset size to the point where it can fit in memory.
For example, binning the detector dimensions of a 256$\times$256$\times$256$\times$256 12-bit dataset by 2 along each axis reduces the size by a factor of 4 from 8.6~GB down to a more manageable 2.1~GB.

\begin{figure*}[hbt!]
  \centering
      \includegraphics[width=14.5cm]{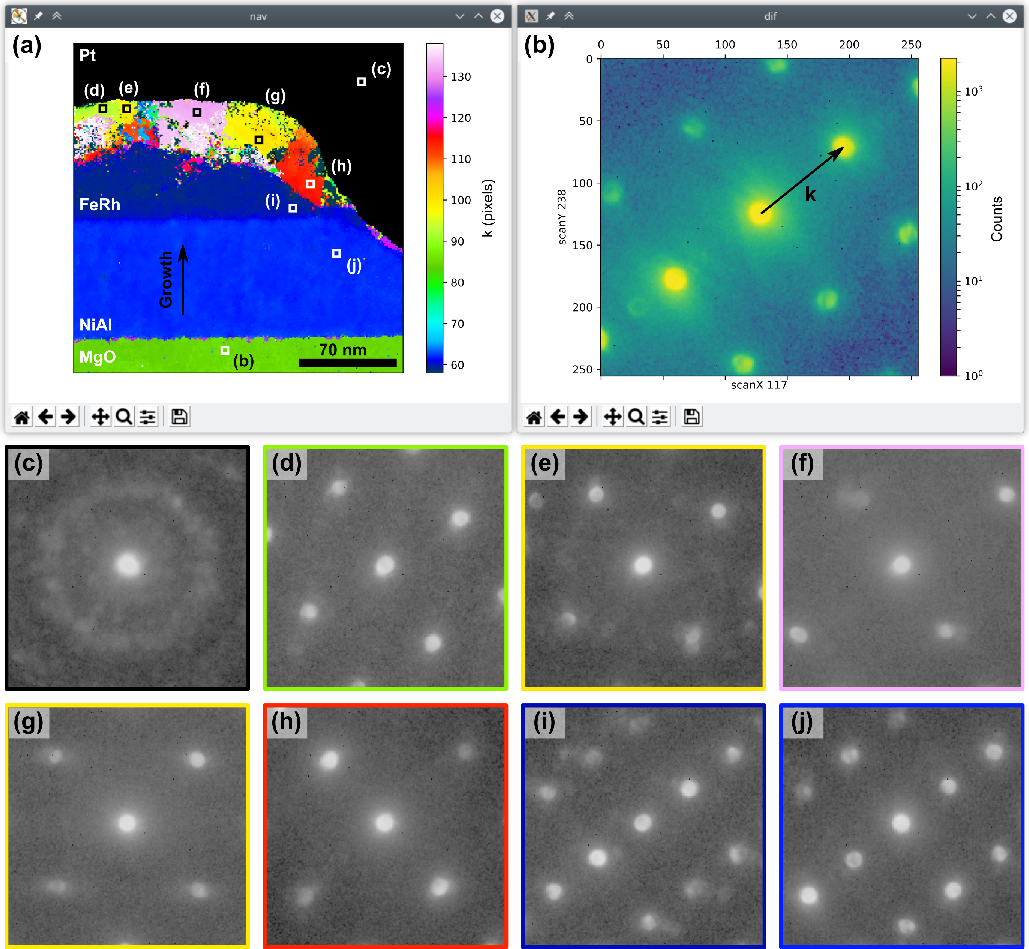}
      \caption{Example of the \texttt{DataBrowser} class showing (a) a user supplied navigation image and (b) the diffraction pattern from the scan point marked by the white square at the bottom of (a). The dataset is from an FeRh sample \cite{Temple_PRM_2018_FeRh, Temple_PRM_2018_FeRh_data} and the navigation image is the diffraction pattern lattice parameter in pixels along the material layer growth direction, obtained by analysing the data using the using the methodology detailed in Section~\ref{sec:lattice}. This direction is marked by the vertical arrow in (a). The equivalent axis is marked by a similar arrow in (b) and is rotated with respect to the scan axes. The black region in (a) is from electron and ion beam deposited platinum and has been masked. All other annotations in (a) have been added to the GUI screenshots to mark the locations of the diffraction patterns shown in panels (c)-(j). Details of the acquisition from the referenced publications are: a scan size of 256$\times$256 points, a probe step size of 0.93~nm, and a semi-convergence angle $<$1 mrad.}
      \label{fig:data_browser}
\end{figure*}

\section{Data Visualisation}
\label{sec:data_visualisation}
Once a dataset has been acquired, it is important to be able to visualise the raw data and to be able to correlate it with the results produced from processing the dataset.
The \texttt{DataBrowser} class of the \texttt{fpd.fpd\_file} module allows basic GUI-based data inspection, using by default the pre-rendered sum image for navigation if one of our HDF5 files \cite{fpd_part1_arxiv} is supplied, and loading images on-the-fly as needed.
The class can also display data from other sources, such as in-memory NumPy arrays and memory mapped files, and so it is also useful when inspecting data immediately after acquisition and before conversion of the data to the HDF5 format.

To demonstrate the \texttt{DataBrowser} class, we use it to display a recently published nanobeam diffraction dataset \cite{Temple_PRM_2018_FeRh_data} from a cross-section of a patterned, epitaxially grown FeRh / NiAl / MgO sample \cite{Temple_PRM_2018_FeRh}.
FeRh, in its B2 structure, exhibits a tunable first-order transition at a temperature of 370~K, which is accompanied by a ferromagnetic to antiferromagnetic ordering transition \cite{Lewis_2016_FeRh_rev}.
These properties make FeRh of interest for a number of applications, including data storage and sensors.
In the following, we present an analysis of the crystal structure, a key property in determining the magnetic ordering, using the methodology detailed later, in Section~\ref{sec:lattice}.

The top row of Fig.~\ref{fig:data_browser} shows screenshots of the two GUI windows of the \texttt{DataBrowser} class.
The navigation image in Fig.~\ref{fig:data_browser}(a) is a user-supplied one showing the diffraction pattern lattice parameter, $k$, normal to the growth direction, as indicated by the black arrows in (a) and (b) (the other annotations in (a) have been added to the GUI image for this presentation of the data).
The diffraction pattern at the location in the scan marked by the bottom white square in (a) is shown in (b).
The displayed diffraction pattern can be selected by clicking anywhere in the navigation image or by dragging the white marker with the mouse, enabling the dataset to be examined.

The extent of the FeRh and NiAl layers and the MgO substrate are readily identifiable from regions of uniform colour in the navigation image of Fig.~\ref{fig:data_browser}(a).
The black region in this image is from nanocrystalline platinum deposited prior to forming the cross-section and has been masked in the image.
Beneath the Pt layer, the top and side sections of the FeRh layer have been modified by a combination of the 1~kV Ar$^+$ etching used in the sample patterning \cite{Temple_PRM_2018_FeRh} and damage done during focused ion beam cross-section preparation, creating disordered regions and grains of different FeRh structures.

While the changes in the mapped lattice parameter are indicative of different grains, the other lattice parameters must be considered when defining a grain.
In Figs.~\ref{fig:data_browser}(c)-\ref{fig:data_browser}(j) we show diffraction patterns outlined in the same colour as the regions they are taken from in Fig.~\ref{fig:data_browser}(a), with the exact single scan pixel location of each pattern marked by the equivalently labelled white or black boxes within each region.
The diffraction patterns from the NiAl layer [Fig.~\ref{fig:data_browser}(j)] and the epitaxial section of the FeRh layer [Fig.~\ref{fig:data_browser}(i)] are both consistent with a B2 structure when viewed along the [011] direction, as reported previously \cite{Temple_PRM_2018_FeRh}, while the MgO pattern [Fig.~\ref{fig:data_browser}(b)] is consistent with the NaCl structure when viewed along the [100] zone axes.
The regions labelled (d)-(h) each have a diffraction pattern consistent with a chemically disordered bcc crystal structure \cite{Temple_PRM_2018_FeRh}, with the rotation of the crystal orientation, lattice parameter, and strain varying between regions (\textit{c.f.} Figs.~\ref{fig:data_browser}(e)-\ref{fig:data_browser}(g)).
Over the imaged area, approximately half of the FeRh material is in a disordered phase.

In addition to the structural analysis of the type discussed above, the ability to inspect the 4-D dataset in this manner is particularly useful in magnetic imaging, where the source, either magnetic or crystalline, of apparent beam-shifts can be interrogated by navigating the 4-D dataset using a colour vector-magnitude image produced from the analysis (examples of vector-magnitude plots of this type may be seen in \citealp{Paterson_2019_PRB_asi}).
This type of beam-shift processing will be discussed further in Part~III.

More configurable plotting of data from our HDF5 files \cite{fpd_part1_arxiv}, such as plotting along different axes, may be achieved by opening the EMD formatted \cite{emd_format} datasets embedded in the HDF5 file in HyperSpy \cite{hyperspy_library}.
This can be performed with the \texttt{fpd\_to\_hyperspy} function of the \texttt{fpd.fpd\_file} module or by loading them directly through the \texttt{pixStem} or HyperSpy libraries.
Furthermore, the datasets may also be loaded into any custom analysis or visualisation code written in Python using the \texttt{fpd\_to\_tuple} function which relies on the h5py library \cite{collette_python_hdf5_2014}, or indeed any of the many other languages supporting the HDF format \cite{hdf5_file_format}.

\section{Virtual Detector Images}
\label{sec:virtual_detectors}
Traditional STEM detectors are routinely used to generate image contrast by collecting the signal from electrons scattered through different angles by the sample under study, from bright-field (BF) \cite{LeBeau_2009_PRB_BF, MacLaren_2015_APLmat_haadf_bf}, through annular bright-field (ABF) \cite{HAMMEL_1995_UltraMicros_ABF, FINDLAY_2010_UltraMicros_ABF} to annular dark-field (ADF), and high angle annular dark-field (HAADF) \cite{Pennycook1991_um, HARTEL_1996_UltraMicros_haadf}.
However, these detectors offer a limited range of often mutually exclusive fixed collection angles at any given camera length.
By using a pixelated detector, the range of scattering angles is resolved and this provides great flexibility in the both the number and range of collection angles from which images may be constructed by applying `virtual' apertures to the dataset in software \cite{Rauch_2005_virt_det, VDF_Rauch_Veron_2014}.
In addition to allowing greater insight into the sample properties, pixelated detectors have the potential benefit of being more dose efficient than segmented diode or photomultiplier tube detectors \cite{Shibata2010_detector} by avoiding repeated scans when multiple collection angles are required, and also do not require the careful calibration \cite{Jones_JoM_2018_stem_det_cal} that traditional STEM detectors do in order to minimise errors in quantitative imaging.
Furthermore, when there are no de-scan coils in a system, or where they are not perfectly set up, movement of the probe position on the detector will occur; an effect that is especially apparent in large area scans.
This can be corrected for in software when using a pixelated detector, whereas with fixed geometry detectors, a range of collection angles depending on the probe position would be sampled, potentially giving rise to non-intuitive or non-quantitatively-interpretable contrast in the resultant image.

In many cases, this effect is not a significant one, but when it is, the data may be corrected during HDF5 file creation~\cite{fpd_part1_arxiv} or the data may be corrected afterwards.
Multiple methods may be used to determine the direct beam position.
Thresholded centre of mass and edge-filtered phase correlation techniques, such as those implemented in the \texttt{fpd.fpd\_processing} module can be used in many cases, and will be discussed in detail in Part~III with regards to differential phase contrast imaging.
Canny edge detection combined with Hough transforms may also be used, but this tends to be much slower than other methods due to the increased computational requirements of the technique.
This approach, implemented in the \texttt{find\_circ\_centre} function of the same module, is demonstrated in Section~\ref{sec:direct_beam} along with an edge fitting technique. 

Accurate knowledge of the centre position is clearly important when using it as basis for making radial profiles in order to prevent mixing of scattering angles and, thus, maintain angular resolution.
All of the aforementioned methods for centre finding work well for diffraction patterns acquired with a low convergence angle beam in cases where there is no overlap between the diffraction discs.
When there is a large degree of overlap between the discs, such as in data acquired with a high convergence angle beam, the task becomes more difficult (examples of these sorts of images are shown in Fig.~\ref{fig:holz_proc}(a-c), discussed later).
If the sample is on-axis, the aforementioned methods should work well.
However, a small misalignment of the zone-axis is a common issue in crystalline TEM samples, due to the sample thinning process imparting a slight bend in the samples.
This can cause the intensity in the centre part of the diffraction pattern to shift, resulting in a misidentification of the centre position of the direct beam.
For some cases, edge detection followed by Hough transforms can work surprisingly well, as can thresholded centre of mass, if the threshold is chosen carefully.
Other ways of mitigating this is scanning a smaller area, this limits the misalignment (and also de-scan), but is not always desirable when large fields of view are required.
A potential solution to this issue in particularly difficult samples is to acquire a reference dataset in vacuum with exactly the same experimental settings, except for the sample stage position.
Here, only the direct beam is visible, so calculating the centre position is trivial, but care must be taken to ensure that the sample does not deflect the beam due to electromagnetic fields, electrostatic charging, or the sample being slightly tapered.
Another possibility is having reference areas within the dataset, for which the centre position can be interpolated for the whole field of view.
Aspects of these issues, in particular, de-scan correction, will be covered more thoroughly in Part~III.

\begin{figure}[hbt!]
  \centering
      \includegraphics[width=8.5cm]{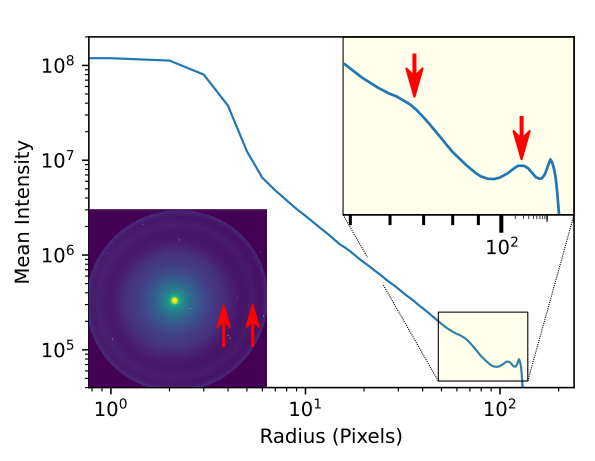}
      \caption{Radial distribution of scattering intensity from the summed diffraction pattern, shown in the inset, of a 256$\times$256 probe position scan of a mouse liver microtome section, calculated using the \texttt{fpd.fpd\_processing.radial\_profile} function. The pixel spacing was 1.2~nm, the exposure was 4~ms, the semi-convergence angle was 0.436~mrad, the camera length was 180~cm, and the condenser aperture was 10~$\muup$m. The arrows mark the location of peaks of intensity.}
      \label{fig:rdf}
\end{figure}

\begin{figure*}[hbt!]
  \centering
      \includegraphics[width=16.0cm]{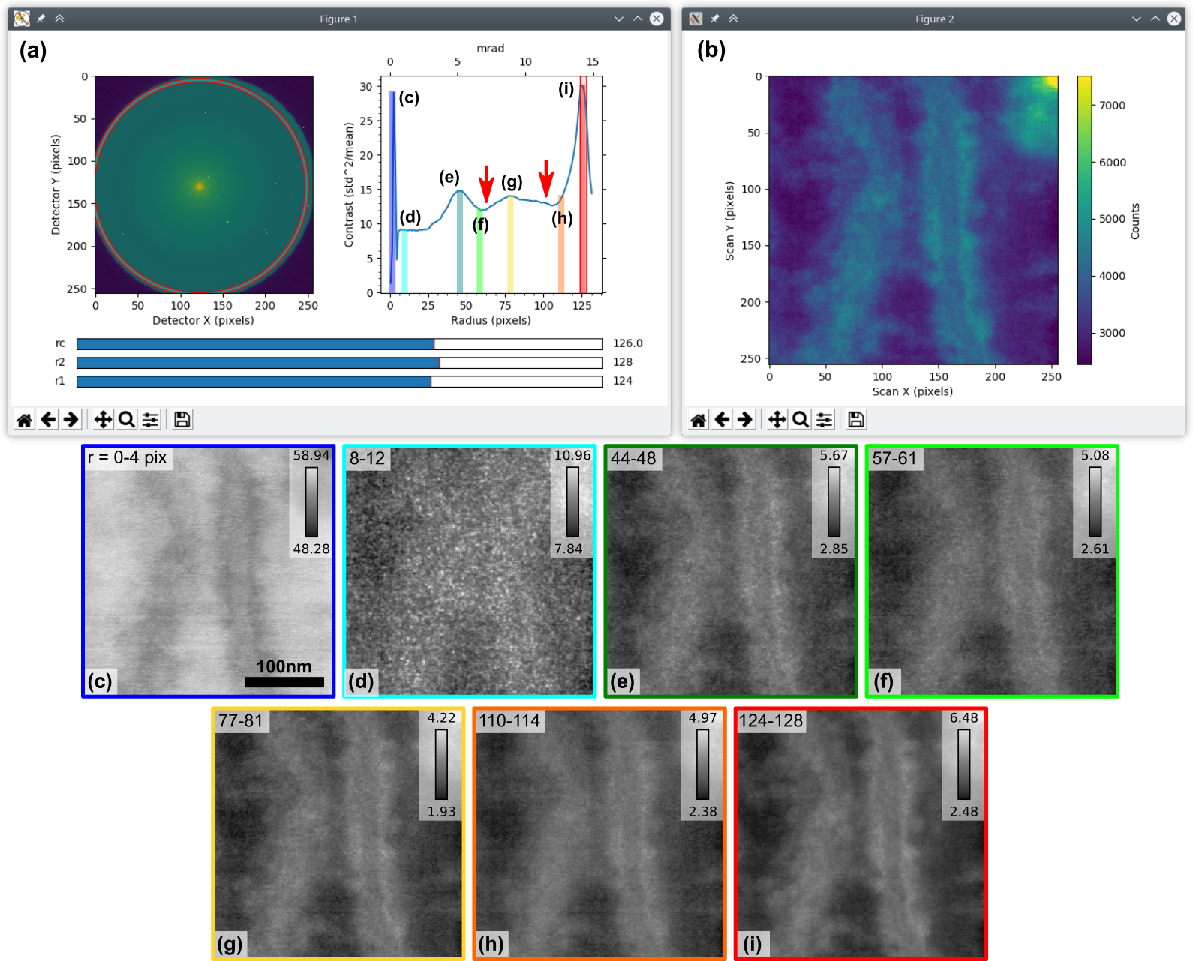}
      \caption{(a), (b) Example of the \texttt{VirtualAnnularImages} class, allowing interactive plotting of virtual annular detector images with instantaneous updates. The sample data is the same as that used for Fig.~\ref{fig:rdf}, a scan of a mouse liver microtome section showing endoplasmic reticula studded with ribosomes. The partial height coloured windows and annotations in (b) have been added to the regular display to mark the locations of the detector geometries used to generate the virtual aperture images in (c)-(i), from bright-field to dark-field. The arrows in (a) show the location of the peaks in the radial distribution presented in Fig.~\ref{fig:rdf}. The intensity ranges in (c)-(i) are set to 0.2\% to 99.8\% of the range for each panel, and the colour bars in each panel show the counts in thousands.}
      \label{fig:VirtualAnnularImages}
\end{figure*}

Some idea about the characteristic scattering angles of a sample may be obtained from direct inspection of a single or averaged diffraction pattern, or by calculating the radial profiles of their intensities.
Depending on the sample properties, this process may be improved by examining other measures of contrast, such as the variance or maximum values across the scan axes.
For the biological sample used here, there is little difference in the apparent contrast from these measures and, as we will show, it is not always simple to determine which scattering angles yield the best contrast in virtual images.
As an example, in Fig.~\ref{fig:rdf} we plot the radial distribution of scattering intensity from a microtome section of a mouse liver.
The inset shows the diffraction pattern summed over all scan positions (this is the same as what is sometimes referred to as a position averaged convergent beam electron diffraction (PACBED) pattern, without dividing by the number of scan points) that was pre-calculated during conversion of the 4-D data to our HDF5 format \cite{fpd_part1_arxiv}.
We will stay in pixels for the simplicity of the discussion, but the images may be calibrated in scattering angle or, equivalently, reciprocal space units (such as k or Q).
Marked by arrows in the diffraction pattern are two rings of different widths; these are more easily visible at around pixels 63 and 102 in the mean intensity profile (also arrowed).
The drop in intensity beyond in the corners of the image marks the aberration-limited field of view of the probe corrected JEOL ARM200cF STEM \cite{mcvitie2015_magtem} in the objective-off mode that was used to record the data.

The scattering angles marked by the peaks in Fig.~\ref{fig:rdf} may be used to define one or more virtual apertures.
These are simply arrays with values between 0 and 1, corresponding to regions outside and inside the collection angles of interest, respectively, and are applied to the whole dataset by multiplication and then summing the total counts for each mask.
Methods for doing this are provided in the \texttt{synthetic\_aperture} and \texttt{synthetic\_images} functions from the \texttt{fpd.fpd\_processing} module, and by the \texttt{virtual\_annular\_dark\_field} method of \texttt{pixStem}'s \texttt{PixelatedSTEM} class.
The resulting virtual detector images allow the contributions to different scattering angles to be spatially resolved.
This approach also works well when using arbitrary, non-annular masks, such as when selecting one or more diffraction spots, where the contrast produced may be used as to determine the location and extent of different phases or lattice defects \cite{Gammer_UM_2015_virt_ap}.

With out-of-core processing from compressed HDF5 files, these calculations may take several seconds, depending on the scan size, the computer hardware, and the data chunking.
This approach is thus well suited to generating images from known collection angles.
While inspection of the diffraction pattern is suitable for identifying the relevant angles in simple cases, it is far more useful to gain live feedback on the changes in spatial contrast from alteration of the virtual detector collection angles.
To enable this, the \texttt{fpd.fpd\_processing} module also provides the \texttt{VirtualAnnularImages} class.
This class maps the \texttt{radial\_profile} function, discussed above, across all scan points using the \texttt{map\_image\_function} function of the same module, and then generates weighted cumulative sums of the radial profiles that may be looked-up at a later point to form the virtual detector image from any given detector geometry.
This intermediate dataset is typically $\sim$40-50 times smaller than the 4-D one and is currently stored within the class, since it can easily fit in memory.
For the data in Fig.~\ref{fig:rdf}, the size is reduced from 8.6~GB to 195~MB.
The significantly smaller size of 3-D datasets also makes them much more amenable to multivariate analyses such as non-negative matrix factorisation (NMF), independent component analysis (ICA), and principal component analysis (PCA).    

Once instantiated, virtual images with any possible continuous set of scattering angles may be generated near-instantaneously using the \texttt{annular\_slice} method of the class.
This calculation time is on the order of milliseconds and, unlike the more flexible \texttt{synthetic\_images} approach, the image calculation time does not change with aperture size.
A current limitation of this efficiency saving is that centre position is taken to be constant across the scan, however it is not a fundamental one and it could be removed in a future version.
As was noted above, in many cases, this limitation is not an issue and, when it is, techniques are available to align the data.

Live feedback on the influence of the virtual detector angles on image contrast may be obtained using the \texttt{plot} method of the \texttt{VirtualAnnularImages} class.
Figure~\ref{fig:VirtualAnnularImages} shows the two plots created by the \texttt{plot} method [Figs.~\ref{fig:VirtualAnnularImages}(a) and \ref{fig:VirtualAnnularImages}(b)] applied to the same data as was used in Fig.~\ref{fig:rdf}, and images created from the dataset [Figs.~\ref{fig:VirtualAnnularImages}(c)-\ref{fig:VirtualAnnularImages}(i)].
The two vertical features are endoplasmic reticula, each studded with ribosomes.
Figure~\ref{fig:VirtualAnnularImages}(a) is a navigation window containing the virtual detector controls, while Fig.~\ref{fig:VirtualAnnularImages}(b) shows the live image produced by the virtual detector.
The navigation window contains a user supplied navigation image (in this case, a summed diffraction pattern), a measure of the real-space image contrast as a function of the scattering angle, and controls for the starting, ending and centre radii for the virtual detector.
The detector position and extent are indicated by the red lines in the two plots, and the radii may be changed via the horizontal slider controls with instant updates to the virtual detector image.
The real-space image contrast is measured here by the ratio of the variance in the virtual detector image, $\sigma^2$, to the image mean, $N$.
For a flat image with Poissonian noise, $\sigma^2$ is purely from the noise and is equal to $N$, so the contrast measure, $\sigma^2 / N$, would give a value of 1.
For non-flat images with the same noise properties, contrast values greater than 1 indicate image \emph{signal}.
Additional noise in the source images will increase the base level and alter the linearity of the contrast parameter of the virtual detector images, but should remain monotonic with image signal.
Similar normalised variance measures are routinely applied in fluctuation electron microscopy \cite{Voyles_2002_UM_fem}, where the variation is calculated across the azimuth at each radii (scattering angle) in each single image \cite{Hart_2016_fem}.
The contrast measure used here is different; it is for the \emph{image} produced from each single pixel width detector, calculated across all radii near-instantaneously using the class itself.

The peaks in the scattering angle dependence plot show which angles produce strong contrast, and are a useful aid in determining the range of angles to focus on.
In the case of the biological sample shown, several peaks are visible, but none of them align with those seen in the radial distribution of Fig.~\ref{fig:rdf} (these are also marked by red arrows in Fig.~\ref{fig:VirtualAnnularImages}(a)), contrary to what one would expect if the local peaks in diffraction intensity produce images with more contrast. 
Figures~\ref{fig:VirtualAnnularImages}(c)-\ref{fig:VirtualAnnularImages}(i) show the virtual images produced at the geometries indicated in the annotations, from bright-field to annular dark-field, at locations between and at the peaks of maximum contrast.
The geometries are also indicated in Fig.~\ref{fig:VirtualAnnularImages}(a) by the partial height windows matching the order of the images and their coloured outlines.

Within the dark-field, there are three peaks of the contrast parameter, at approximately 46, 79 and 126 pixels [Figs.~\ref{fig:VirtualAnnularImages}(e), \ref{fig:VirtualAnnularImages}(g) and \ref{fig:VirtualAnnularImages}(i)], and these are where the images do indeed show the highest contrast and range of spatial frequencies, allowing the ribosomes to be better resolved, with the best contrast produced from the highest of these scattering angles.
To estimate the SNR of each individual image, we apply to them the single image autocorrelation method \cite{Thong_2001_1imageSNR} implemented in the \texttt{snr\_single\_image} function of the \texttt{fpd.utils} module.
This approach exploits the fact that noise is spatially uncorrelated, whereas the signal \emph{is} correlated, allowing extrapolation of the autocorrelation function to zero image displacement to be used to estimate the noise-free signal power and, thus, the noise power.
The optimum extrapolation method depends of the properties of the image and our implementation supports a number of methods.
For the data in Fig.~\ref{fig:VirtualAnnularImages}, a linear extrapolation method is adequate and gives values of 20-37 for the first six images [Figs.~\ref{fig:VirtualAnnularImages}(c)-\ref{fig:VirtualAnnularImages}(h)], and 71 for the highest scattering angle image [Fig.~\ref{fig:VirtualAnnularImages}(i)], confirming that it does indeed have a much higher SNR.
These SNR values are the power values with the signal magnitudes taken from the changes in the images, as defined in the original method, and thus give measures of the useful contrast.

For the biological sample used here, there is very little Bragg scattering and the ADF contrast changes in the angular range acquired of up to $\sim$15~mrad is likely to be due to incoherent Rutherford scattering from light elements.
The rings visible in the diffraction pattern and its radial distribution are equivalent to d-spacings of approximately 0.35 and 0.22~nm, and may be indicative of characteristic bond lengths or periodicities in molecules but, as we have shown, do not always yield images of the greatest useful contrast.

The ability to interrogate and reconstruct STEM images showing different contrast from a single scan, post-acquisition, as outlined above, makes the technique of virtual detectors a very powerful one.
In addition, unlike the similar hollow cone technique \cite{KRAKOW_1976_hollow_cone, Tsai2016_hollow_cone}, only one scan is required to generate many types of contrast, making it dose efficient, which is especially useful in beam sensitive materials.

While we demonstrate the utility of applying virtual detectors with data from a biological sample, the technique is useful in many other applications, where different navigation images may be more appropriate.
For example, in polycrystalline samples, the navigation image may be a diffraction pattern from a specific grain, identified from navigating the dataset with the \texttt{DataBrowser} class (see Fig.~\ref{fig:data_browser}) or a similar method.
It is then straightforward to set annular apertures that select specific spots, thereby identifying the extent of the grain of interest and the locations of similar grains.
More advanced analysis for out-of-plane and in-plane structural analysis is discussed next, in Sections~\ref{sec:holz} and \ref{sec:lattice}, respectively.

\section{Higher Order Laue Zone Analysis}
\label{sec:holz}

\begin{figure*}[hbt!]
  \centering
      \includegraphics[width=15.0cm]{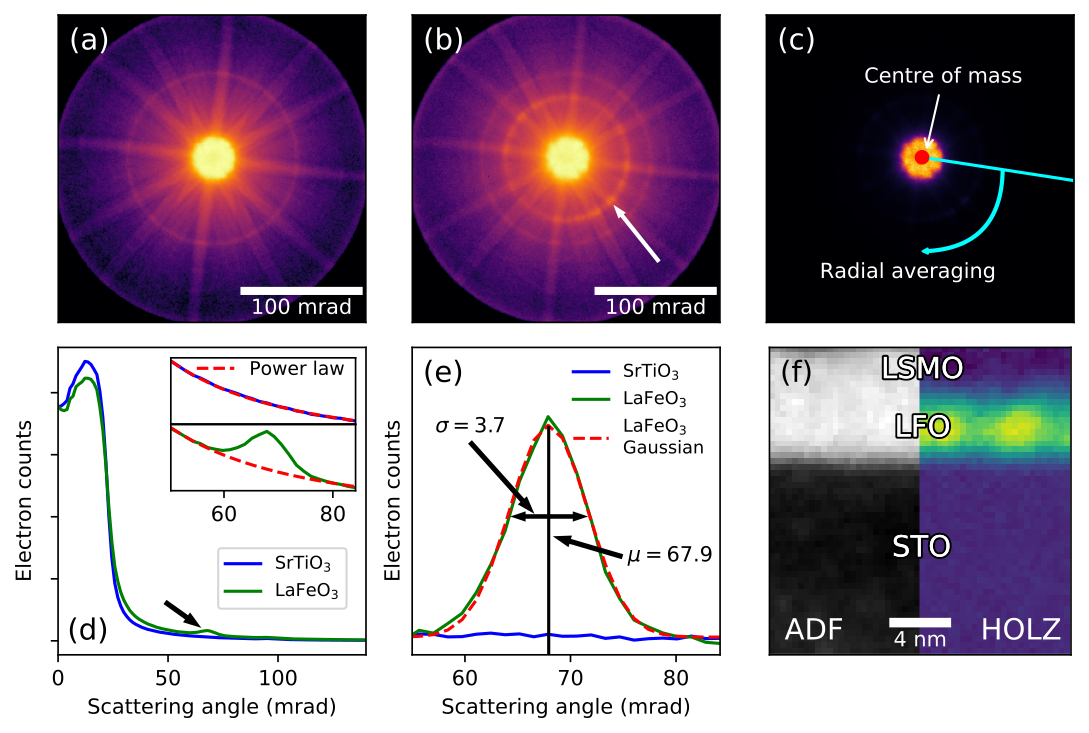}
      \caption{Processing of higher order Laue zone (HOLZ) diffraction rings from a La$_{0.7}$Sr$_{0.3}$MnO$_3$ (LSMO) and LaFeO$_3$ (LFO) bilayer film grown on SrTiO$_3$ along the (111) direction.
      The data was acquired on a probe corrected JEOL ARM200cF, with an acceleration voltage of 200~kV, a convergence semi-angle of 20.4~mrad, a scan step size of 0.37~nm and with a total scan size of 64$\times$64 points.
      STEM diffraction pattern of the (a) STO and (b) LFO, plotted on a logarithmic scale, with the electron beam parallel to the (110) direction.
      LaFeO$_3$ has an extra inner HOLZ ring due to doubling of the unit cell (arrowed).
      (c) Schematic of the processing, first using thresholded centre of mass to find the centre of the pattern, then radial averaging across the azimuth around this point.
      (d) Radial distribution of (a) and (b).
      The arrow shows the inner HOLZ peak from (b).
      The insets to (d) show the region around the HOLZ peak, with a power law fitted to the background.
      (e) Region around the HOLZ peak, with the power law from (d) subtracted, showing the background is accurately removed, and with a 1-D Gaussian profile fitted to the data. Numbers showing the centre position and sigma of the Gaussian.
      (f) Result of the processing for the bilayer system, showing the bright LSMO and LFO regions in the virtual annular dark field (ADF, 106-163 mrad) image, and bright LFO region in the HOLZ contrast.}
      \label{fig:holz_proc}
\end{figure*}

The intersection of the Ewald sphere with parallel reciprocal planes of a crystal gives rise to concentric circles of reciprocal space where the Bragg condition is met, leading to constructive interference in the diffraction pattern.
The central region of the diffraction pattern is formed by the crystal structure \emph{perpendicular} to the electron beam, and is called the zero'th order Laue zone (ZOLZ).
Diffraction spots in higher order Laue zones (HOLZ) correspond to intersection of Ewald's sphere with parallel planes of reciprocal lattice points offset \emph{along} the electron trajectory (formally the electron wavevector in reciprocal space) \cite{Emslie_ElectronDiffraction_1934, HOLZ_Jones_Rackham_Steeds_1977}.
Thus, for an on-axis diffraction pattern, periodicity parallel to the beam results in rings of intensity in the diffraction pattern, as shown in Figs.~\ref{fig:holz_proc}(a) and \ref{fig:holz_proc}(b).
The radius, $r$, of the ring in reciprocal space for the first-order Laue zone is given by:
\begin{equation}
      r = 2\sin^{-1}\left(\sqrt{\frac{\lambda}{2 d_z}}\right)
\end{equation}
where $\lambda$ is the wavelength of the electron, and $d_z$ is the lattice period parallel to the electron beam.
An equivalent relation was originally derived by \citealp{Emslie_ElectronDiffraction_1934} in a different form that incorporated the camera length.
Similar relations may be derived for higher order Laue zone rings.
It has long been recognised that high angle coherently-scattered intensity concentrated in HOLZ rings may affect the contrast in HAADF images \cite{SpenceZuo_HOLZ_1989}.

Because of the dependence of the radius on the out-of-plane structure, the presence of HOLZ rings can be used to obtain information about the structure parallel to the electron beam: the smaller the radius of the Laue zone ring, the larger is the distance between atomic planes parallel to the electron beam. 
This was originally proposed as a method for studying altered periodicity in dislocation core structures \cite{SpenceKoch_HOLZ_2001} and was later successfully used to determine the periodicity along the beam direction in sodium cobaltite using a simple thin annular detector setup \cite{Huang_holz_stem_2010}.

Alternative approaches to obtaining such information from crystals are to tilt the sample to high angles, or take multiple specimens containing the same kind of feature cut at different angles and reconstruct using either diffraction tomography \cite{Kolb_2007_um_sped_tomo, Mugnaioli2009_um_sped_tomo} or atomic resolution discrete tomography \cite{MacLarenRichter_2009, MacLaren_2015_APLmat_haadf_bf, Van_Aert_DiscreteTomo_2011}.
These approaches are not applicable to all samples (e.g. they could not be used on a dislocation core).
They are also time-consuming and either use statistical reconstruction from multiple specimens or multiple sample areas (and so assume that the different areas imaged all contain exactly the same structure, just imaged along different directions), or, in the case of the tomographic approach, use a large radiation dose on one area.
Moreover, using a thin annular detector is inflexible and needs camera length tuning to get this to work, thus using a fast pixelated detector and detecting and processing the HOLZ data after acquisition is far more flexible for a variety of systems.

One example of a class of materials with structural distortions that cause a change in the size of the unit cell is perovskite oxides, which often exhibit a doubling of the unit cell as a result of octahedral tilting \cite{Glazer_TiltingPerovskites_1972}.
These distortions are important to characterise, as they can heavily influence functional properties such as ferromagnetism and ferroelectricity.
This is especially important in thin film systems, as much of the interesting physics resides in the detailed crystallographic structure of these films.
One such example is La$_{0.7}$Sr$_{0.3}$MnO$_3$ (LSMO) and LaFeO$_3$ (LFO) bilayer films grown on SrTiO$_3$ (STO) \cite{ingrid_lsmo_lfo_sto_2016, Nord_2018_holz}.
Whilst, previous work has been performed using annular bright field or bright field imaging in STEM \cite{Aso2013_scirep_octahedral, Nord2017_atomap, Wang2016_um_oxygen_octahedra_picker, Kim2017_um_oxygen_octahedral} to characterise the in-plane oxygen atom movements associated with the octahedral tilting, it is also possible that there are atom movements resulting in a periodicity change along the beam direction.
Specifically, whilst researchers in perovskites and related oxides often talk about octahedral tilting, this rarely happens in isolation and is usually associated with the modulation of cation positions, a feature that is likely to result in a significant diffraction signal.

Figure~\ref{fig:holz_proc}(a) shows the STEM diffraction pattern of the non-distorted cubic STO substrate imaged along the (110) direction, which has one visible Laue zone ring.
The diffraction pattern of the LFO film is shown in Fig.~\ref{fig:holz_proc}(b), which is non-cubic due to structural distortions, and is similar to the LSMO pattern except for having an extra Laue zone ring at lower scattering angles (arrowed).
The intensity of this additional ring is proportional to the amount of distortion and can, for example, be used to characterise the strength of the atomic movements as a function of position in thin film structures \cite{Nord_2018_holz} as described below.
Additionally fine structure can appear in the HOLZ rings and they can split into more than one ring, which may reveal additional subtle details about the structure \cite{Peng_ACA_1989_holz}, and this can be recorded in scanned diffraction datasets at a suitable camera length.

The first step in the processing [Fig.~\ref{fig:holz_proc}(c)] is finding the centre position of the diffraction pattern for each probe position.
This is necessary due to the fact that the shift and the tilt of the electron beam is not always perfectly separated in the scanning coils, especially for larger scan areas, and there is no de-scan coil set up in the microscope used in this work, resulting in the diffraction pattern centre moving as a function of beam position.
As discussed in Section~\ref{sec:virtual_detectors}, finding the centre position can be non-trival for high convergence angle diffraction patterns like we have here.
However, for these samples, the zone-axis remained sufficiently aligned across the scan area for it to be possible to find accurate centre positions using a thresholded and masked centre of mass procedure.
The mask removes the higher scattering angles, so only the region containing the bright-field disc is included.
Then, the thresholding sets all values below the chosen threshold to zero, and all above it to one.
The centre of mass calculation is done on this masked and thresholded image, resulting in the centre position of the diffraction pattern.
For more difficult sample data, acquiring a reference dataset in vacuum and performing the same centre of mass calculation on this dataset can often provide accurate centre positions.
Some alternatives techniques for centre finding are demonstrated in Section~\ref{sec:direct_beam}.

Next, integration is performed along the azimuth for each probe position to generate radial dependencies, as previously discussed in Section~\ref{sec:virtual_detectors}, using the aforementioned centre positions of the diffraction patterns.
This reduces the 4-D dataset to 3 dimensions: 2 probe dimensions and 1 dimension with intensity as function of scattering angle, as shown in Fig.~\ref{fig:holz_proc}(d).
By removing one dimension, the data size is reduced by a factor of around the number of pixels on the longest axis of the detector (256, in our case), making it much more manageable and easier to fit into the computer memory.
This radial averaging is also useful for data exploration of large datasets like this, due to the much smaller data size, as demonstrated in Section~\ref{sec:virtual_detectors} on virtual detectors.
One complication is distortions caused by the projection system, leading to features which should be round becoming elliptical.
When doing the radial averaging, this distortion causes a broadening of the HOLZ peak (Fig.~\ref{fig:holz_proc}(d,e)).
However, for sufficiently small distortions, this does not heavily affect the total intensity in the peak.
Ideally, this should be corrected for by calculating the diffraction distortions from a reference dataset.
However, as the focus of this analysis was the \textit{intensity} of the peak, no such corrections were performed here.

After the radial distribution has been calculated, the HOLZ ring is reduced to a HOLZ peak shown by the arrow in Fig.~\ref{fig:holz_proc}(d).
To extract the intensity from only the peak, a power law is simultaneously fitted to both the regions before and after the peak.
The fits are shown in the insets in Fig.~\ref{fig:holz_proc}(d), both for the STO which does not have the extra HOLZ peak, and for LFO which has an extra HOLZ peak at lower scattering angles.
While a power law is not the optimal function to fit to this type of background over long distances, it works well for extracting the HOLZ peaks over a relatively short range of scattering angles, as shown by the level background of the background-corrected profiles in Fig.~\ref{fig:holz_proc}(e).

After background-correction, the profiles may be analysed by simply summing the intensity in the HOLZ peak to give a map of the level of structural distortion, as shown in Fig.~\ref{fig:holz_proc}(f).
Pairing the HOLZ processing with virtual annular dark field generated from the same radially averaged data, one can see that the extra HOLZ ring is indeed present only in the LFO layer.
More information can also be extracted by fitting a 1-D Gaussian to the peak, as shown in Fig.~\ref{fig:holz_proc}(e), giving information about changes in the scattering angle (centre position) and variation in scattering angle (standard deviation).
Further information on the material aspects from a detailed analysis of this system are published elsewhere \cite{Nord_2018_holz}.
This type of analysis is best suited for monocrystalline materials, especially epitaxial thin films and heterostructures, or polycrystalline materials with large grains (e.g. domain-structured functional oxides with different orientations of the principal axes of the cell), as the technique requires the crystal to be imaged along a zone axis.
Whilst it may be performed at atomic resolution, this is not necessary, and this technique has the advantage that it yields information about the modulation of atomic positions along a column without requiring atomic resolution, as demonstrated in previous analyses of this effect \cite{Borisevich_AtomicShape_2010, Azough_TB_2016}.

\section{Lattice Analysis}
\label{sec:lattice}

\begin{figure*}[hbt!]
  \centering
      \includegraphics[width=16.5cm]{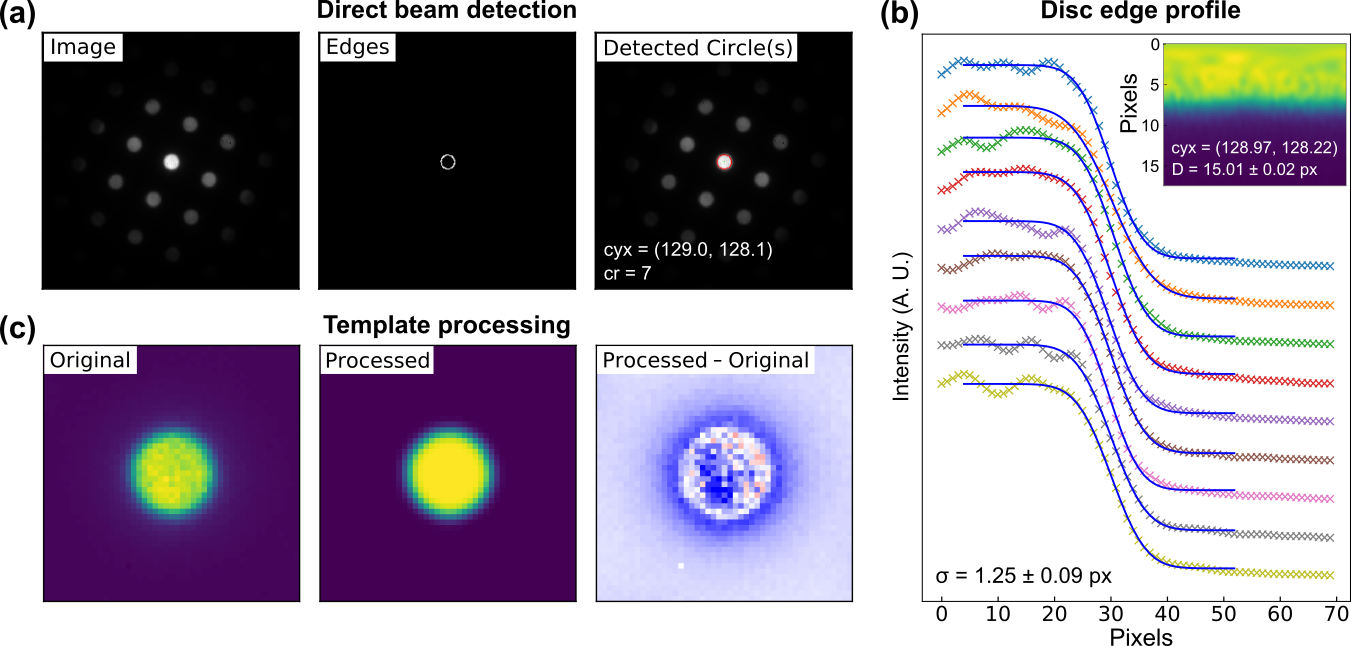}
      \caption{Lattice analysis stage 1: (a) finding the direct beam, (b) characterisation of the disc edge properties, and (c) forming a template using, respectively, the \texttt{find\_circ\_centre}, \texttt{disc\_edge\_properties}, and \texttt{make\_ref\_im} functions from the \texttt{fpd.fpd\_processing} module. The lines in (b) are error functions fitted to the data (symbols) taken from the direct beam image converted to polar coordinates, plotted in the inset. The inset also shows the measured diameter of the disc, D, and the optimised centre coordinates of the extracted edges. The error in the latter is 0.01 pixels.}
      \label{fig:sped_step1}
\end{figure*}

Analysing the crystal structure of materials has been a part of electron microscopy since its inception.
A number of techniques in transmission electron microscopy can be used for atomic resolution imaging, most frequently phase contrast TEM (high resolution TEM, HRTEM) and HAADF STEM.
Nevertheless, diffraction-based techniques have been the main method for structure solution or analysis throughout the history of electron microscopy, notwithstanding the advantages of the image-based techniques for giving local crystallographic information, especially where this varies with position such as in thin films, heterostructures, and around defects and precipitates.

SAED can localise the sample region from which data is collected to $\sim$100s~nm, much higher than the nanometre or even {\AA}ngstr{\"o}m length scales that imaging with high energy electrons can potentially allow.
Recently, however, scanned diffraction techniques have been on the increase, especially with the rise of faster electronic detectors (initially CCD/scintillator arrangements, but more recently direct electron detectors). 
Moreover, the development of scanning precession electron diffraction (SPED) \cite{Vincent_1994_UM_sped} has allowed collection of spatially resolved diffraction data down to areas of 2~nm or less with very high precision of the reciprocal lattice parameters.

Convergent beam techniques provide the highest spatial resolution by imaging with a probe formed by a large diffraction-limiting aperture, and it is these techniques that have seen the greatest application in materials science in a number of areas, including grain size and orientation determination, molecular structure solutions, and material strain measurement.
Strain is a particularly important property that influences the functionality of a wide range of materials, with perhaps the most notable class of materials being semiconductor devices \cite{Cooper_2016_semi_strain_micron, Bashir_JAP_2019_sped_Ge}.

A number of techniques for strain measurements have been developed \cite{Beche2013_um_strain_comparison}, including dark-field electron holography (DFEH) \cite{Hytch2008_dfeh, Cooper_2009_apl_dfeh, Beche_2011_dfeh}, NBED \cite{nbed_beche_2009}, SPED (also referred to as nanobeam precession electron diffraction (NPED)) \cite{Midgley_IUCrJ_2015_ped_rev, Rouviere_APL_2013_SPED}, scanning moir\'{e} fringe (SMF) analysis \cite{Su2010_um_SMF, Naden2018_moire_stem}, HRTEM geometrical phase analysis (GPA) \cite{Hytch1998_gpa}, and atomic column spacing displacement characterisation \cite{Nord2017_atomap}.
Of these, the best fractional strain precision reported is with SPED at 2$\times$10$^{-4}$ \cite{Rouviere_APL_2013_SPED}, however, very recent work with patterned probes \cite{Zeltmann_UM_2020_patterned_aps, Guzzinati_2019_bessel_arxiv} have achieved precisions approaching the same value.

Analysis of such 2-D lattice images can be performed with a range of packages including Atomap, optimised for atomic resolution STEM imaging \cite{Nord2017_atomap}; CrysTBox for HRTEM, selected area electron diffraction (SAED), and CBED imaging \cite{Klinger17_crystbox}; library based approaches to crystal phase and orientation identification \cite{Rauch_SPED_2010}; and ones recently developed specifically for 4D-STEM \cite{py4dstem, Zeltmann_UM_2020_patterned_aps}.
With the rise in the use of fast pixelated detectors in TEM and STEM and the consequent ability to record large data volumes in short times, the need arises for the ability to analyse large numbers of diffraction patterns accurately and automatically.
These include images with point-like lattice vertices, created by imaging under Fraunhofer conditions, and in convergent beam electron diffraction (CBED) imaging, where the lattice vertices are discs.
In the following sections, we report simple processing methodologies that are applicable to general lattice analysis and, in particular, to diffraction patterns produced by techniques including CTEM, CBED, NBED, and SPED.

To demonstrate our technique, we apply it to a commercially available crystalline MgO substrate imaged in a JEOL ARM200cF TEM operated at 200~kV in SPED mode using a custom \cite{MacLaren_MandM_2020_sped} DigiSTAR precession system from NanoMEGAS.
MgO is a widely studied material \cite{Yang_MSEC_2006_MgO} that is used in magnetic tunnel junctions \cite{Parkin_Nature_2004_MgO, ZHU_Materials_today_2006_MTJ} and is of interest as a room temperature ferromagnetic insulator when strained \cite{Zhenghe_JAP_2015_MgO}.
Unstrained MgO has a cubic NaCl crystal which produces a diffraction pattern with a square projection of the reciprocal space lattice when viewed along the $<$100$>$ zone axes.
The sample was prepared by application of a standard focused ion beam milling procedure \cite{Schaffer2012_um_fib_sample_prep} to a multilayered structure.
Electron energy loss spectroscopy measurements in probe-corrected STEM mode with convergence and collection semi-angles of 29 and 36~mrad, respectively, using a GIF Gatan QuantumER 965 spectrometer, showed that the ratio of thickness, $t$, to inelastic mean free path, $\lambda$, ($t/\lambda$) of the sample was 0.59.
Using an MgO density of 3.58~g/cm$^2$ \cite{Egerton_EELS}, $\lambda$ is 137~nm, giving the thickness of our MgO sample as 81~nm, sufficiently thick for there to be significant dynamical effects from multiple elastic scattering.

The standard precession system was modified \cite{MacLaren_MandM_2020_sped} to employ a Medipix3 detector instead of a fast video recording of the fluorescent screen using an external CCD camera.
This modification enables improved imaging fidelity and efficiency through the higher detective quantum efficiency and the noise-free readout properties of the direct electron counting detector.
Compared to static probes, precessing the electron beam reduces dynamical effects by incoherently averaging over a range of diffraction conditions to give a pseudo-kinematical diffraction pattern, resulting in more uniform diffraction pattern discs \cite{Midgley_IUCrJ_2015_ped_rev}.

In the experiment, test data was acquired by imaging MgO along a [100] axis under the following conditions: the beam was precessed at an angle of 1.5$^\mathrm{\circ}$ over a 10~ms exposure, the microscope was operated in TEM-L mode, with a spot size of 5, the condenser aperture was 20~$\muup$m, the semi-convergence angle was determined from the MgO lattice to be 1.1~mrad, the camera length was 100~cm, and the scan pixel size was 2.2~nm.
No exploration of the acquisition parameter space or subsequent optimisation of the parameters was performed.

\begin{figure*}[hbt!]
  \centering
      \includegraphics[width=16.5cm]{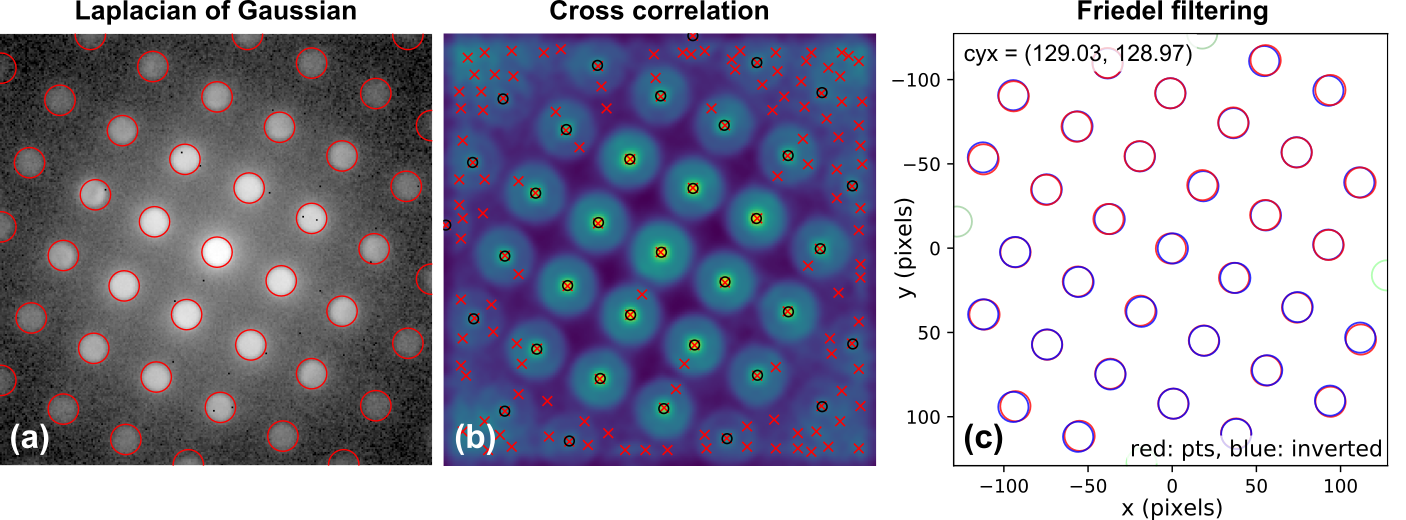}
      \caption{Lattice analysis stage 2: extracting features and filtering them by inversion symmetry using the \texttt{blob\_log\_detect} and \texttt{friedel\_filter} functions of the \texttt{fpd.tem\_tools} module. The red circles in (a) show the blobs detected using Laplacian of Gaussian (LoG) applied to a logarithmically scaled image. (b) The results of edge-filtered cross-correlation with a template image, with red crosses indicating peak locations and black circles those filtered for the next step by the LoG data. Green circles in (c) show the discs removed by `Friedel' filtering (see main text for details).}
      \label{fig:sped_step2}
\end{figure*}

Correcting for the overestimation of the electron counts recorded by the Medipix3 by a factor of 4 due to scattering of 200~kV primary electrons between pixels (see \citealp{McMullan2007_mpx2_electron_imaging} for a discussion), we estimate the dose to be approximately 4.5$\times$10$^5$~e$^-$/nm$^2$.
While this is above the low dose regime needed for most soft or beam sensitive materials of 10$^2$-10$^5$~e$^-$/nm$^2$ \cite{Yakovlev_Micron_2008_dose_EELS}, there is great scope to reduce the dose through reduction in beam current or, as pointed out by \citealp{MacLaren_MandM_2020_sped}, by increasing the precession rate in other microscopes that support it.

To enable the SPED datasets to be used more easily, the \texttt{topspin\_app5\_to\_hdf5} function of the \texttt{fpd.fpd\_io} module allows conversion of data originally recorded in the flat native NanoMEGAS TopSpin app5 format to the multidimensional HDF5 format outlined in Part~I \cite{fpd_part1_arxiv}.
Alternatively, the Merlin system \cite{Plackett_2013_merlin} through which the Medipix3 data is acquired can be programmed to output the data directly to a raw file whilst the acquisition is being performed and controlled by the TopSpin software.
These datasets can be processed by the \texttt{fpd} and \texttt{pixStem} libraries in a number of ways, including the previously covered virtual detector imaging and Laue zone analyses, and differential phase contrast (DPC) analyses for field mapping, which will be covered Part~III.
Indeed, the benefits of SPED brought about by averaging over a range of diffraction conditions has great potential to also improve DPC imaging, as others have very recently also noted and investigated \cite{Mawson_2020_sped_DPC}.

Machine learning approaches to structure analysis have recently been applied to SPED data \cite{Martineau_ASCI_2019_sped_machine}.
In contrast, the methodology described below is a `bottom up' one that optimises precision without knowledge of the context of each image.
The processing methodology is modular in design, allowing modification or addition of processing steps, as needed for the specific application and, indeed, could be complemented by machine learning approaches.
Below, we outline the main steps without going into the many options provided to tailor the analysis to the sample data and, subsequently, assess the precision of the technique applied to this dataset.

\begin{figure*}[hbt!]
  \centering
      \includegraphics[width=15cm]{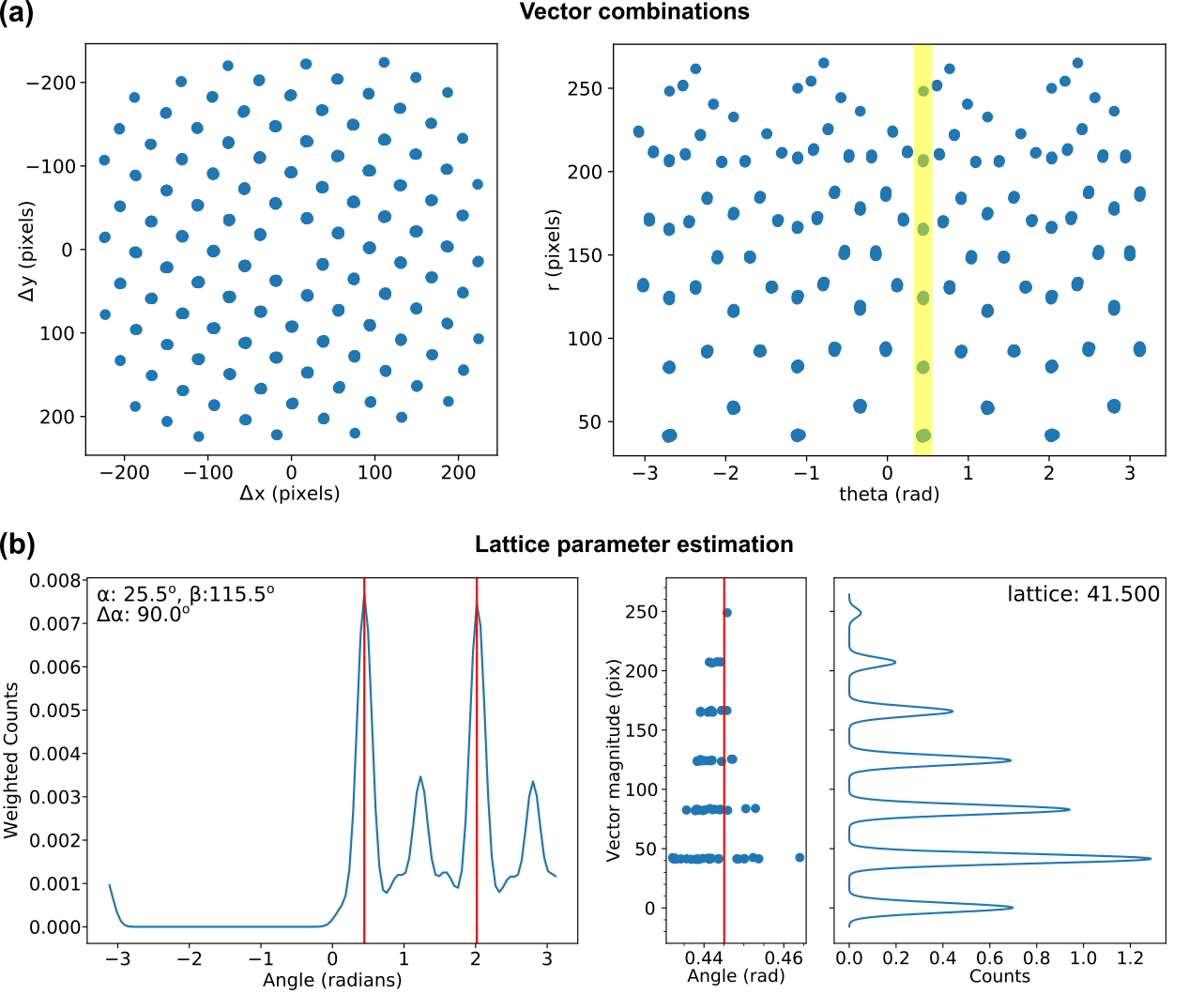}
      \caption{Lattice analysis stage 3: lattice parameter estimation. (a) Lattice vector combinations in Cartesian (left) and polar (right) coordinates. (b) Lattice angle (left) and magnitude (right) estimates (only one of the two lattice vector magnitude plots is shown).}
      \label{fig:sped_step3}
\end{figure*}

\begin{figure*}[hbt]
  \centering
      \includegraphics[width=14cm]{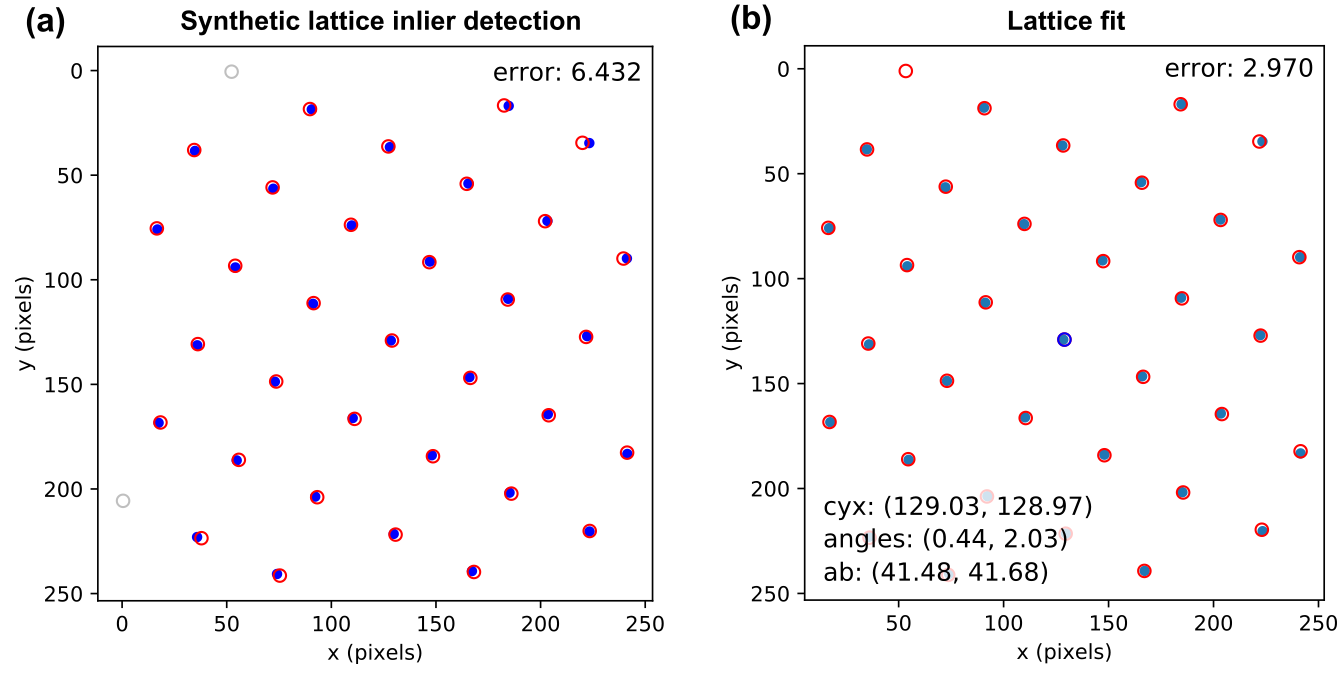}
      \caption{Lattice analysis stage 4: lattice parameter extraction. (a) Analysis data (solid symbols) filtered by synthetic lattice (open symbols) inlier detection, and (b) the lattice optimised through least squares fitting, where the total error between the two lattices (indicated in the annotations) in minimised. The blue symbol in (b) shows the direct beam location. Outliers would be marked by crosses, however, all data are inliers in this case.}
      \label{fig:sped_step4}
\end{figure*}

The general process can be divided into the four main stages of (i) direct beam detection and characterisation, (ii) feature extraction and filtering, (iii) lattice parameter estimation, and (iv) synthetic lattice inlier detection and fitting.
These steps are outlined in Figs.~\ref{fig:sped_step1}, \ref{fig:sped_step2}, \ref{fig:sped_step3}, and \ref{fig:sped_step4}, which include slightly modified versions of the the plots optionally produced by the relevant analysis functions applied to the first diffraction pattern of the MgO SPED scan.
Each figure will be discussed in turn; the final results of the analysis are shown in Fig.~\ref{fig:sped_results} and will be discussed later.

\subsection{Direct Beam Characterisation}
\label{sec:direct_beam}
The first stage in the process is to find the direct beam position and size.
Figure~\ref{fig:sped_step1}(a) shows three images produced in this analysis.
The first shows the original image, the second shows the image processed with Canny edge detection, and the third shows the results of a circular Hough transform \cite{Gouillart_2016_python_scikit}, where the red circle represents the position and size of the detected disc.
This transform generates a 2-D Hough space for each radius, with values proportional to the correlation of the edge image and the nominal circle centred at each possible coordinate (the dimensions of the space).
The centre of the optimised circle is obtained with sub-pixel resolution by fitting 2-D Gaussians to each peak in Hough space, or the images may be upscaled.
If the lattice vertices are point-like, then the size of the circle represents an effective radius.
If the vertices are discs, such as those produced by a convergent beam (or are of any other shape), then a template of the beam image may optionally be formed for use in edge-filtered cross-correlation in order to estimate the position of all vertices in the next steps.
As several others have noted, procedures of this type can improve the accuracy of the image registration results by relying more on feature shapes than image intensities \cite{Schaffer_UM_2004_alignment, Pekin_UM_2017_disc_accuracy, matus_pixelated_stem_magnetic_2016}.
An important parameter in the analysis is the disc edge profile, and this can be estimated by converting the central beam image to polar coordinates, as shown in the inset in Fig.~\ref{fig:sped_step1}(b), and then fitting error functions to the edge region, as shown in the main panel.
Through this process, the disc diameter and a new centre position are also obtained from fitting to the angle dependence of the extracted error curve centre positions. 
Finally, a filtered reference image may be created from the central beam image, using the extracted disc edge width, as shown in Fig.~\ref{fig:sped_step1}(c).
This step further reduces the influence of intensity diffracted from the bright-field disc, which will not be uniform across all scan points and within all diffraction discs.
These non-uniformities in intensity are most easily visible as texture in the rightmost image in Fig.~\ref{fig:sped_step1}(c), which shows the difference between the real disc and the uniform processed one.

This first stage need only be done once for each dataset, unless the direct beam moves a significant distance during the scan.
If this is the case, the direct beam position may be estimated at each scan position using the technique described above, by centre of mass as discussed in Section~\ref{sec:holz}, or by any other means.

\subsection{Vertex Identification}
The second stage in the analysis is to find the position and size of all the potential lattice vertices and then to filter these so that only those with inversion symmetry are retained.
The results of one way of processing the images to extract potential lattice points is shown in Fig.~\ref{fig:sped_step2}.
The first step is to perform Laplacian of Gaussian (LoG) blob detection with a supplied radius range, based on the direct beam radius (or effective radius for Fraunhofer imaging conditions).
The detected features are shown in Fig.~\ref{fig:sped_step2}(a) by the red circles.
If the lattice vertices are point-like, 2-D Gaussians may optionally be fitted to extract sub-pixel peak locations.
For non-point-like or disc-shaped vertices, pixel-resolution edge-filtered cross-correlation (CC) can be used with the template image of the direct beam produced in the first stage.
This result is shown in Fig.~\ref{fig:sped_step2}(b), where the CC peaks, marked by red crosses, indicate the location of maximum correlation, corresponding to the identified vertex locations.
By the nature of lattices, cross-correlation can give rise to many false peaks, as visible in the figure.
Therefore, these are filtered by the LoG data which is in general more reliable but which has a higher noise level.
Specifically, only those points located within some distance of the LoG vertex coordinates are kept.
The peaks retained in this step are marked by black circles.
Also at this stage, 2-D Gaussian functions can be optionally fitted to the vertices to extract the peak locations with sub-pixel accuracy, as was done here.
A comparison of the different analysis levels will be given in Section~\ref{sec:lattice_analysis_precision}, with reference to Fig.~\ref{fig:sped_results_comp}.

Next, the data is filtered according to Friedel's law \cite{Friedel_1913}, removing points that, within some relative and absolute error, have no equivalent point at the location inverted about the centre coordinate.
Most of the potential vertices are retained in this example, as shown in Fig.~\ref{fig:sped_step2}(c), with only those at the edges being removed (shown in green).
In this step, the centre coordinate may be updated based on the systematic differences in positions of the pairs of diffraction spots, which allows small changes in the position of the direct beam to be accounted for.
In this case, the $(c_y, c_x)$ value was updated to (129.03, 128.97), as shown in the figure annotations.

\subsection{Lattice Estimation}
The previous two stages only extract the positions and sizes of potential lattice vertices and assume nothing of the relationship between these positions, the lattice properties.
In the next two stages, the lattice properties are estimated and then optimised from the filtered vertex locations, as shown in Figs.~\ref{fig:sped_step3} and \ref{fig:sped_step4}, respectively.

There are many potential ways of estimating lattice parameters and we discuss two of these.
First, we use a simple statistical approach, as shown in Fig.~\ref{fig:sped_step3}.
To improve the statistics, all combinations of lattice vectors are computed [left of Fig.~\ref{fig:sped_step3}(a)] and converted into polar coordinates [right of Fig.~\ref{fig:sped_step3}(a)].
Next, a histogram is generated from the polar data [left of Fig.~\ref{fig:sped_step3}(b)], with optional 1/r$^2$ weighting applied to increase the significance of nearer neighbour data, where $r$ is the Euclidean distance.
An expected symmetry of the lattice may be provided at this stage to further improve the analysis.
Two peaks are identified from this histogram, marked by the red lines, corresponding to two lattice vector angles.
The polar data is then sliced at these angles and a histogram generated [right of Fig.~\ref{fig:sped_step3}(b)]; the equivalent location is marked by a yellow band in the right hand panel of Fig.~\ref{fig:sped_step3}(a).
The histogram is processed using peak detection or Fourier analysis to extract the lattice magnitudes along these directions, yielding a complete estimate of the lattice parameters.

This simple approach to lattice estimation works best when the sample is on or near a low order zone axis and there is a single crystallographic phase within the beam diameter.
When multiple lattices are present within a single image, alternative methods to lattice parameter estimation such as clustering and inlier detection may prove to be useful, and can easily be incorporated in the presented processing methodology without modification of the previous or next steps.

The \texttt{lattice\_from\_inlier} function of the \texttt{fpd.tem\_tools} module provides an alternative method of lattice estimation.
This function generates lattices for all combinations of two of each of the first $n$ potential lattice vertex coordinates in distance from the direct beam position, and returns the lattice parameters with the maximum number of inliers to the synthetic lattice, as determined by comparison of the euclidean distance between matched vertices to a supplied threshold.
When multiple lattices have the same number of inliers, one is selected by a user supplied criterion which, by default, is set to the maximum of the geometrical mean of the lattice parameter magnitudes.
This reduces the chances of integer divisions of the lattice parameters being found.
As a consequence of this approach, this function typically returns different equivalent lattice parameters across a uniform material.
The parameters may be homogenised using the \texttt{lattice\_resolver} function described below.
Compared to the statistical approach used above with reference to Fig.~\ref{fig:sped_step3}, this approach is in general more robust against additional spots being present in the image, less robust against there being many missing spots, and produces a slightly less precise but still perfectly usable initial estimate of the lattice.

\begin{figure*}[hbt!]
  \centering
      \includegraphics[width=16.5cm]{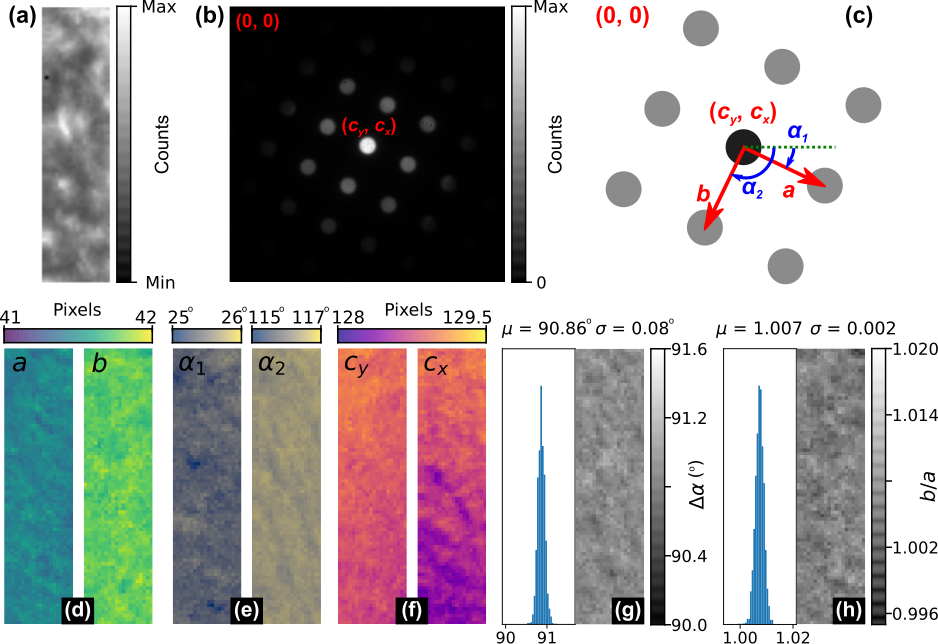}
      \caption{MgO SPED data analysis results. (a) Real and (b) reciprocal space sum images for an 80$\times$20 pixel scan (176$\times$44~nm, pixel size: 2.2~nm) using a 256$\times$256 pixel Medipix3 detector. The reciprocal space coordinate system is shown in the annotations in (b) and in the schematic in (c). (d)-(f) Spatially resolved lattice parameters and, (g) lattice angle delta and (h) lattice parameter magnitude ratio histograms and maps.}
      \label{fig:sped_results}
\end{figure*}

\subsection{Lattice Optimisation}
In the fourth and final step, using an estimate of the lattice magnitudes and angles (generated by any method), a synthetic lattice is created and data inliers to the model are selected.
This synthetic lattice and experimental data are shown as solid and open symbols in Fig.~\ref{fig:sped_step4}(a).
Next, the lattice is fitted to the inliers using least squares optimisation, resulting in the final lattice shown in Fig.~\ref{fig:sped_step4}(b).
Constraints and bounds on and between lattice parameters can be applied during the fitting process, but none were used for this test data.

The modular design of the lattice analysis makes it is easy to assess each processing step, allowing optimisation of the analysis parameters, and to customise the processing by inserting additional steps or removing existing ones.
Once the processing conditions are established on a test image or images, the analysis may be applied to many images in parallel by passing a user created function to the \texttt{map\_image\_function} function of the \texttt{fpd.fpd\_processing} module.

The four stages outlined above often result in a single unit cell per material.
However, as mentioned earlier, it is possible for the particular unit cell returned within a material to change between a number of equivalent unit cells.
To condition the data, as an additional step, the lattice parameters extracted can be resolved in specific directions using the \texttt{lattice\_resolver} function of the \texttt{fpd.tem\_tools} module.
This generates a small synthetic lattice using the extracted lattice parameters and then identifies the new lattice vectors by finding those lattice points at angles closest to user specified values.
This is especially useful when mapping lattice parameters across epitaxial interfaces between differing materials, where in- and out of- plane strain may be of particular interest.
Potential alternatives that are likely to meet with success include clustering and machine learning approaches applied to the extracted lattice parameters or parameters derived therefrom.
Compared to applying ML approaches to images directly, applying them to the extracted data would vastly reduce the size of the data to be processed (by a factor of several thousands), and also automatically accounts for any de-scan in the measurement as the calculated basis vectors are insensitive to the pattern centre position.

\subsection{MgO Results}
\begin{figure}[hbt!]
  \centering
      \includegraphics[width=8.5cm]{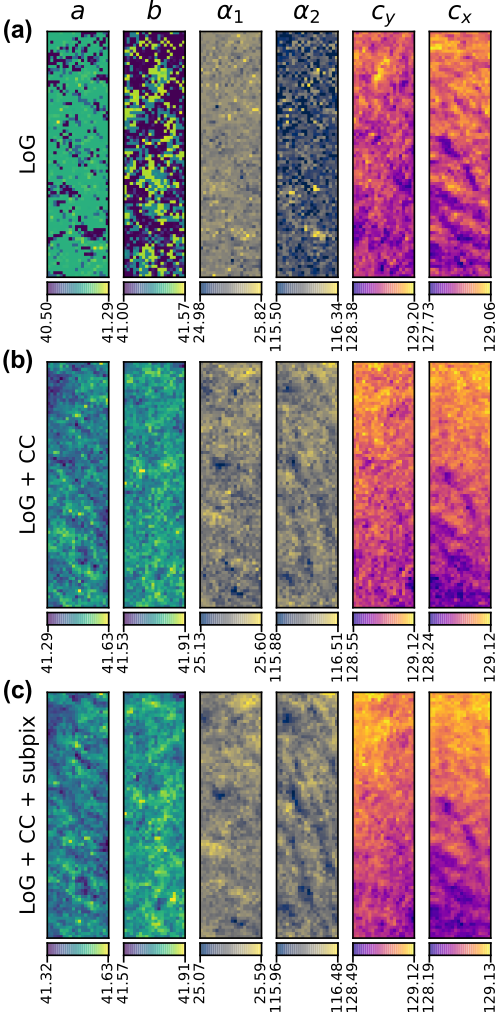}
      \caption{Comparison of analysis of the MgO SPED data used in Fig.~\ref{fig:sped_results} with three different approaches (in rows) of generating the diffraction disc centres: (a) Laplacian of Gaussian (LoG), (b) LoG plus cross-correlation (CC), and (c) LoG plus CC plus sub-pixel peak fitting. The units of the data are the same as in Fig.~\ref{fig:sped_results}, pixels and degrees. The colour bar ranges are matched to the data range of each panel.}
      \label{fig:sped_results_comp}
\end{figure}

Figure~\ref{fig:sped_results} shows the results of analysing the complete MgO SPED dataset, using a reference image for cross-correlation and applying sub-pixel peak fitting.
A comparison of the different levels of processing applied the same data is shown in Fig.~\ref{fig:sped_results_comp}, and will be discussed in Section~\ref{sec:lattice_analysis_precision}.

Figures~\ref{fig:sped_results}(a) and \ref{fig:sped_results}(b) show the real and reciprocal sum images for the dataset.
The reciprocal space coordinate system is defined in Figs.~\ref{fig:sped_results}(c).
The origin is located at the top left corner, the coordinates of the direct beam are ($c_y$, $c_x$), and the lattice parameters are defined by the magnitudes, $a$ and $b$, and the angles, $\alpha_1$ and $\alpha_2$.
The six lattice parameters extracted from the data are shown in Figs.~\ref{fig:sped_results}(d)-\ref{fig:sped_results}(f), with the angles between lattice vectors, $\Delta \alpha$, and the ratio of the lattice magnitudes, $b/a$, shown in Figs.~\ref{fig:sped_results}(g) and \ref{fig:sped_results}(h), respectively.

The mean angle between lattice vectors in the SPED data is 90.86$\pm$0.08$^\mathrm{\circ}$ (0.09~\%), and the mean ratio of the lattice vector magnitudes is 1.007$\pm$0.002 (0.16~\%), making for a slightly skewed lattice, and is evidence of strain in our multilayer sample.
By comparison, the equivalent lattice vector magnitude ratio for the nanodiffraction MgO data in Fig.~\ref{fig:data_browser} from a different sample and analysed by the same method, was 1.020$\pm$0.006 (0.6~\%).
Although both datasets are influenced by non-uniformities in the MgO, the difference in the variance of the data will reflect to some extent the improvements obtainable by precessing the electron beam.

Care in interpreting the results must be taken in order to rule out the influence of image distortion \cite{knut_muller_nbed_2019}.
This may be done in a number of ways, including: recording two or more scans at different sample rotations; recording a reference lattice against which strain parameters \cite{Rouviere_UM_2005_gpa} can be calculated; or by correcting the lattice vertices by the application of an affine or other transformation to the vertex coordinates from a suitable calibration.

The centre coordinates approximately describe planes due to de-scan (as discussed before).
However, in samples supporting magnetic or electric fields or exhibiting mean inner potential changes, the beam shifts extracted in these parameters can give useful DPC contrast.
Aspects of DPC analysis will be discussed in Part~III of this work.

\subsection{Analysis Precision}
\label{sec:lattice_analysis_precision}
Figure~\ref{fig:sped_results_comp} shows how the cumulative addition of edge filtered cross-correlation [Fig.~\ref{fig:sped_results_comp}(b)] and sub-pixel peak finding [Fig.~\ref{fig:sped_results_comp}(c)] each improve upon the LoG analysis [Fig.~\ref{fig:sped_results_comp}(a)].
The third row contains the same data as shown in Figs.~\ref{fig:sped_results}(d)-\ref{fig:sped_results}(f), but the colour map scale here is maximised for each panel, with the limits of the ranges shown on each colour bar.

The same main features are present in the LoG (first row) data as are in the third row data, but the noise level is much higher and there is a degree of digitisation, obscuring the more subtle features.
Edge-filtered cross-correlation (CC) gives a large improvement, with the spread in parameter values generally reduced and the results are less digitised.
This is because the processing technique reduces the influence of intensity variations within the discs and because it is inherently more sensitive to the location of the discs.
This will be especially true when the beam is not precessed, such as in CBED or NBED patterns where non-uniformities in the disc intensities can be significantly larger.
The improvement over the LoG plus CC results [Fig.~\ref{fig:sped_results_comp}(b)] obtained by the addition of peak fitting, shown in Fig.~\ref{fig:sped_results_comp}(c), are mainly in the deep sub-pixel range, with a reduction in the random noise level.

To estimate the signal-to-noise ratio in these analyses, we applied the same single image autocorrelation power SNR estimation \cite{Thong_2001_1imageSNR} methodology used in Section~\ref{sec:virtual_detectors} to the data shown in Fig.~\ref{fig:sped_results_comp}.
Here, a second order polynomial extrapolation of the autocorrelation function was used after de-trending the images with a plane fit, and the signal level was taken from the known zero rather than using the variation in the images.
The resulting power SNR values for the $a$ parameters are 64~dB and 70~dB for the LoG plus CC and LoG plus CC with sub-pixel peak fitting analyses, respectively (the results from the other parameters are similar).
The pixelation in the LoG data means we cannot easily extract an accurate SNR estimate for this dataset using this method.
However, an estimate of the SNR may be made using the LoG plus CC with sub-pixel peak fitting analysis data as a reference, and this yields a value of 47~dB for the LoG analysis.

The smooth, continuously varying features of our test MgO data means we cannot interpret the level of precision of our approach from the standard deviation of the parameters.
However, because we have extracted an estimate of the SNR ratio, we may use this parameter to estimate the fractional precision as $1/\sqrt{\mathrm{SNR}}$.
The results give a fractional precision of 3.1$\times 10^{-4}$ (0.03~\%) in the LoG plus CC with sub-pixel peak fitting analysis which, with an $a$ parameter of 41.5 pixels, corresponds to a fractional precision of 0.01 pixels of the detector.
The fractional precision of the LoG plus CC analysis is about half as good at 6.6$\times 10^{-4}$.
The theoretical full width half maximum (FWHM) spatial resolution of the Airy probe in our measurements, estimated using the \texttt{airy\_fwhm} function of the \texttt{fpd.tem\_tools} module, is 1.1~nm, which is consistent with direct imaging of the probe.
However, at the precession angle used in the experiment (1.5$\mathrm{^o}$) and for our sample thickness (81~nm), the centre of the probe will describe circles of a diameter of 2.1~nm at the surfaces of the sample and, thus, the actual spatial resolution of the measurement will be poorer than that defined by the static probe.
In fact, ignoring the increased weighting of the sampling of the specimen in its vertical centre, the FWHM defined resolution of the probe will be given by the diameter of the circle described by the beam.
Importantly, if our 2.2~nm scan pixel data were oversampled, our precision estimates may be overly optimistic.
As a simple check of this, we repeated the analysis using only every other pixel, giving an interpixel spacing of 4.4~nm, and found the SNR values of the LoG plus CC and LoG plus CC with sub-pixel peak fitting analyses decreased to 62~dB and 64~dB, corresponding to precision values of 8.2$\times 10^{-4}$ (0.08~\%) and 6.0$\times 10^{-4}$ (0.06~\%), respectively.
Using the alternative approach of wavelet-based Gaussian noise estimation \cite{Donoho_1994_wavelet_noise_est, Gouillart_2016_python_scikit} yielded very similar fractional precision results to those from the previous analysis with a larger inter-pixel spacing: 7.9$\times 10^{-4}$ for the LoG plus CC analysis, and 5.6$\times 10^{-4}$ for the  LoG plus CC with sub-pixel peak fitting approach, giving credence to the results.
Rebinning the data by a factor of 2 along each axis before performing the SNR calculations, gives precision values around 4.5$\times 10^{-4}$ for both methods.
This improvement over using every other pixel is partly due to the increased dose at the same spatial resolution.

Further work is required to determine the optimum acquisition parameters for the accuracy of the analysis.
For example, the noise-free readout of the Medipix3 and the relatively low number pixel count of 256$\times$256 for a single die (\emph{v.s.} several 1000s for a typical CCD detector), may mean that better results would be obtained by increasing the camera length so that only the lowest order spots are imaged.
Alternatively, different DEDs with higher pixel counts may be used, or more pixels may be added to a Medipix3 detector by tiling the detector chips.
The Medipix3 family of detectors are 3-side buttable \cite{Ballabriga_2013_Medipix3RX}, and thus may be tiled in 2$\times$N arrays \cite{Bucker_2020_nat_commun_mpx_tile_struct}, with larger pixels at the joints which must be accounted for. 
Furthermore, employment of the through-silicon via feature of the Medipix3 family can allow tiling on all four sides of the sensor in even larger arrays \cite{Tick_2011_JoI_tsv, Ponchut_2015_JoI_tsv_smartpix} with minimal dead areas for the readout circuitry.

At the small level of material distortion found here, the fractional precision of the lattice parameters is the same as that of the associated strain parameters.
Very recent reports of strain measurements using patterned probes in the literature \cite{Guzzinati_2019_bessel_arxiv, Zeltmann_UM_2020_patterned_aps} approach the best values reported from standard Airy probes in SPED acquisitions \cite{Rouviere_APL_2013_SPED}, but with potential benefits of improved dose efficiency.
The fractional precision of 6$\times 10^{-4}$ obtained here with a DED is approximately 3$\times$ higher than the value from the latter (of $\leq$2$\times$10$^{-4}$) with a similar spatial resolution, but using exposures 100$\times$ smaller and a detector with 64$\times$ fewer pixels (256$\times$256 DED \textit{vs.} a 2k$\times$2k CCD).
Our results correspond to an approximately 3$\times$ higher sub-pixel precision.
We cannot compare the beam dose between the two experiments, but we note that the aperture used in our experiment was 2-4$\times$ the radius of that used in the referenced work which, taking account of the different exposure times, would give a 25-12.5$\times$ reduction in dose in our experiment with a DED \emph{if} the emission currents were the same.
With the noise-free readout of the Medipix3 and the ability to operate in continuous mode, where every individual electron is counted with no gaps, there is great scope for further reducing the sample dose in SPED for use with beam sensitive materials \cite{MacLaren_MandM_2020_sped}.

\subsection{Method Applicability and Efficiency}
\begin{figure}[hbt!]
  \centering
      \includegraphics[width=8.5cm]{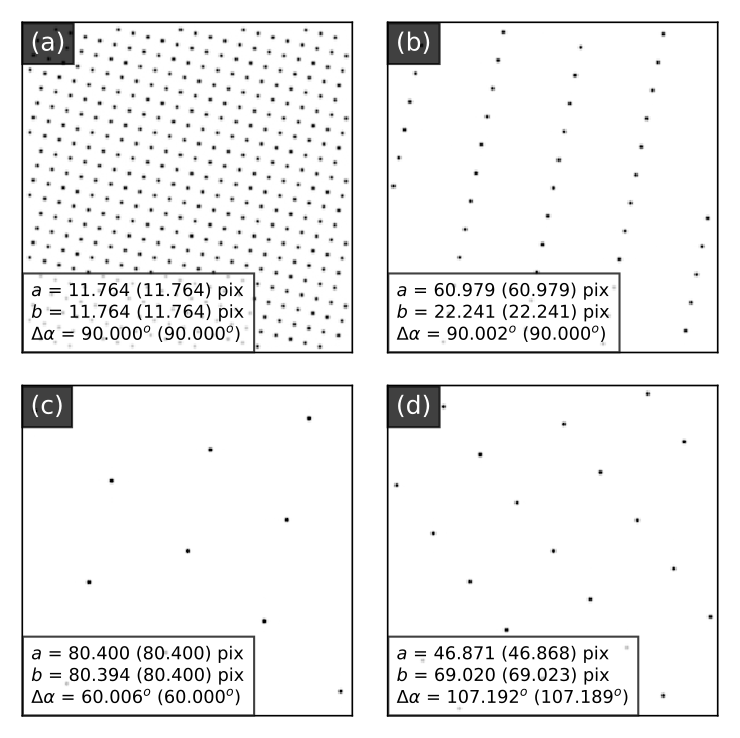}
      \caption{Four different synthetic lattices and, in the inset text, the results of their analysis using identical processing parameters. The nominal values are given in parentheses. All spots were of equal counts before application of Poissonian noise.}
      \label{fig:lattice_shapes}
\end{figure}

We demonstrate the flexibility of our simple lattice analysis method by applying it to the four synthetic lattice images shown in Fig.~\ref{fig:lattice_shapes}.
These images were generated by sub-pixel Fourier shifting 2-D Gaussian spots of peak intensity 1024 and include Poissonian noise.
The lattices have (a) square, (b) rectangular, (c) hexagonal and (d) oblique unit cells.
These cover four of the five shapes available in two dimensional space; the remaining one being the rhombic lattice which only differs from the oblique one by having equal lattice vectors.
The extracted parameters are shown in the insets to Fig.~\ref{fig:lattice_shapes}, with the nominal values in parentheses.
The agreement between the numbers in this somewhat idealised case is less than around 0.006~pixels and 0.006$^\mathrm{\circ}$, corresponding to an average error of 0.003~\%, and gives some idea of the accuracy of the algorithm at this electron dose.
Ultimately, the final noise and precision achievable will depend on the sample; the imaging conditions, including the dose and beam shape; the characteristics of the detector and its number of pixels; and the parameters used in the analysis.

Despite the widely varying lattice size, aspect ratios, and symmetries, the \emph{same} analysis parameters were used to extract the lattice properties across all four datasets in Fig.~\ref{fig:lattice_shapes} at the same time, demonstrating the ability of the approach to characterise multiple materials in a dataset \emph{without} prior knowledge or constraints on what they are.
An example of the application of this feature is given in Section~\ref{sec:data_visualisation}, where multiple phases were identified from experimental nanobeam diffraction data without modification of the analysis parameters.
However, many analysis parameters may be specified to improve the parameter extraction in real data, including the ability to apply arbitrary constraints on the lattice parameters, potentially allowing more accurate data extraction for known materials by removal of unneeded degrees of freedom. 

\begin{figure}[hbt!]
  \centering
      \includegraphics[width=8.5cm]{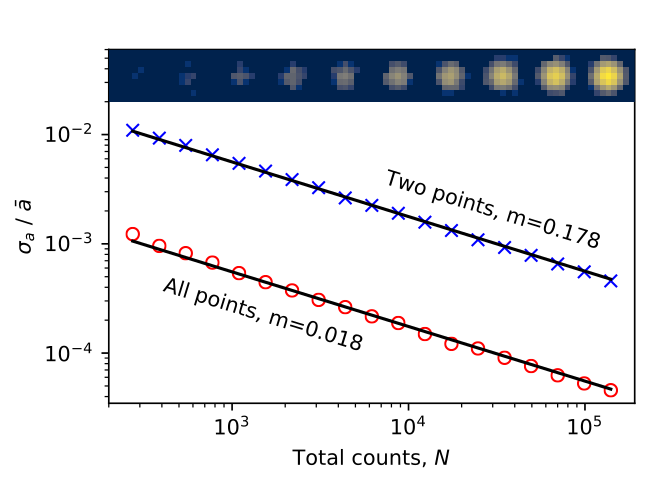}
      \caption{Relative error of the smaller lattice parameter magnitude, $a$, of the lattice in Fig.~\ref{fig:lattice_shapes}(d) as a function of dose. The data was analysed using LoG feature detection plus sub-pixel peak fitting, using the full lattice (`all points') and the central spot and one other point (`two points'). The inset shows example images of every other spot as a function of dose, on a logarithmic intensity scale.}
      \label{fig:lattice_dose_accuracy}
\end{figure}

In data with only Poissonian noise, the relative error of the extracted parameters, $\sigma_v / \bar{v}$, will generally follow
\begin{equation}
  \label{p_err_eqtn}
  \frac{\sigma_v}{\bar{v}} = \frac{m}{\sqrt{N}},
\end{equation}
where $N$ is the total counts and $m$ is a constant that varies with the analysis method used and the properties of the source data, such as the intensity distribution within each lattice spot or disc.
To estimate $m$ in our approach, we calculated the ratio of the standard deviation to the mean of the smaller lattice parameter, $a$ of the synthetic lattice in Fig.~\ref{fig:lattice_shapes}(d) across approximately 1000 images with different Poisson noise, as a function of the total dose, $N$.
This data is plotted as red circles in Fig.~\ref{fig:lattice_dose_accuracy} for the LoG plus sub-pixel peak fitting analysis of each image, along with a fit to the data of the above equation (black line).
The value of $m$ is 0.018 across the dose range investigated of 273 to 140k counts (see inset for examples of a typical lattice vertex as a function of dose).
By comparison, the same accuracy data for an analysis by fitting a 2-D Gaussian function to each of just two of the lattice vertices is shown as blue crosses and has a much higher $m$ value of 0.178.
The equivalent $m$ value obtained by reducing $N$ to the counts in 2 of the 21 vertices is 0.039, around twice that found by using all vertices, showing the benefit of using all available counts.

This approach of fitting to the complete lattice makes use of every vertex and all available signal, optimising the accuracy for a given dose, as well as extracting more information in the form of the other lattice parameters.
The imperfections in real data will mean that the precision of our approach applied to such data will most likely be significantly lower than the characteristic values reported in Fig.~\ref{fig:lattice_dose_accuracy}, and will be dependent on the properties of the sample and the detector, but the benefits of our approach will remain; specifically, that by using all vertices, the signal is used optimally and the effect of systematic variations in vertex properties, such as background variations, will be reduced.

Regardless of the source of the data, once the shape of a lattice has been characterised, it is then possible to extract additional information.
For diffraction data, this includes strain mapping, the number and spatial extent of different phases or grains (see Fig.~\ref{fig:data_browser}), and the spot intensities for structure factor characterisation \cite{Midgley_IUCrJ_2015_ped_rev} and structure determination \cite{Mugnaioli2009_um_sped_tomo, Clabbers_2017_ACD_mpx_tile_struct, Bucker_2020_nat_commun_mpx_tile_struct}, and for the analysis of non-centrosymmetric crystals.
Intensity mapping may be done by generating a synthetic aperture based on the extracted lattice parameters and then applying it to the data, as discussed in Section~\ref{sec:virtual_detectors}, or, for point-like spots, fitting a 2-D Gaussian to the peak regions using the functions provided in the \texttt{fpd.utils} module, which are also used in the sub-pixel peak finding described above, yielding details of the peak properties.

\section{Summary}
In this work, we discussed the key issues around post-acquisition analysis of data from fast pixelated detectors and presented software libraries to allow efficient data processing and visualisation.
We provided examples of their use across a number of applications in the area of structural characterisation, including the techniques of virtual detector imaging for bright and dark-field imaging, higher order Laue zone analysis for extraction of structural information along the path of the beam, and nanobeam and scanning precession electron diffraction for lattice parameter determination and strain analysis.

While the data analysis algorithms and libraries presented are applicable to data from any detector, the examples provided show that highly dose efficient active pixel direct electron detectors such as the Medipix3 perform excellently as universal STEM detectors, despite their relatively low pixel counts.
Indeed, we have demonstrated nanoscale lattice parameter mapping in SPED mode with a fractional precision $\le 6\times10^{-4}$, approaching the best values reported in the literature.
Furthermore, in addition to being of use as a regular STEM detector, with every electron being recorded noise-free, there is excellent prospects for the application of the detector to the characterisation of beam sensitive materials.

The software packages presented are hosted in public repositories \cite{fpd, fpd_demos, pixstem}, are under active development and contain many more features than are covered in this part or in Part~1 \cite{fpd_part1_arxiv} of this work.
These packages are provided under an open source licence, allowing full transparency of the algorithms implemented and for them to be continually improved upon by the community.

Part~III of this work will cover post-acquisition processing and visualisation of data from fast pixelated detectors for differential phase contrast imaging.

\section*{Acknowledgements and author contributions}
G.W.P.and M.N. were the principal authors of the \texttt{fpd} and {pixStem} libraries reported herein (details of all contributions are documented in the repositories), and have made all of these available under open source licence GPLv3 for the benefit of the community. 
R.W.H.W., A.R. and K.A.P. have also made contributions to the source codes in these libraries.
G.W.P and M.N. have led the drafting of this manuscript.
The performance of this work was mainly supported by Engineering and Physical Sciences Research Council (EPSRC) of the UK via the project ``Fast Pixel Detectors: a paradigm shift in STEM imaging'' (grant No. EP/M009963/1). 
G.W.P. received additional support from the EPSRC under grant No. EP/M024423/1. 
M.N. received additional support for this work from the European Union’s Horizon 2020 research and innovation programme under the Marie Skłodowska-Curie grant agreement No 838001.

R.W.H.W., A.R., K.A.P., T.A.M., D.McG. and I.M. have all contributed either through acquisition and analysis of data or through participation in the revision of the manuscript.
The studentships of R.W.H.W. and T.A.M. were supported by the EPSRC Doctoral Training Partnership grant No. EP/N509668/1.
I.M. and D.McG. were supported by EPSRC grant No. EP/M009963/1.
The studentship of K.A.P. was funded entirely by the UK Science and Technology Facilities Council (STFC) Industrial CASE studentship ``Next2 TEM Detection'' (No. ST/P002471/1) with Quantum Detectors Ltd. as the industrial partner.
As an inventor of intellectual property related to the MERLIN detector hardware, D.McG. is a beneficiary of the license agreement between University of Glasgow and Quantum Detectors Ltd.

We thank Diamond Quantum Detectors Ltd. for Medipix3 detector support; 
Dr. Bruno Humbel from Okinawa Institute of Science and Technology and Dr. Caroline Kizilyaprak from the University of Lausanne for providing the liver sample;
Dr. Ingrid Hallsteinsen and Prof. Thomas Tybell from the Norwegian University of Science and Technology (NTNU) for providing the La$_{0.7}$Sr$_{0.3}$MnO$_3$/LaFeO$_3$/SrTiO$_3$ sample shown in Fig.~\ref{fig:holz_proc};
and NanoMEGAS for the loan of the DigiSTAR precession system and TopSpin acquisition software.
Development of the integration of TopSpin with the Merlin readout of the Medipix3 camera has been performed with the aid of financial assistance from the EPSRC under grant No. EP/R511705/1 and through direct collaboration between NanoMEGAS and Quantum Detectors Ltd.

\section*{Data and Software Availability}
The source data and scripts to analyse the data and produce the results presented here are publicly available at \url{https://doi.org/10.5281/zenodo.3903517} \cite{paper_dataset_p2}.
The software presented in this work is available in public Git repositories \cite{fpd, fpd_demos, pixstem}.

\bibliographystyle{mm_nat}

\begin{thebibliography}{107}
\providecommand{\natexlab}[1]{#1}
\providecommand{\url}[1]{\texttt{#1}}
\expandafter\ifx\csname urlstyle\endcsname\relax
  \providecommand{\doi}[1]{doi: #1}\else
  \providecommand{\doi}{doi: \begingroup \urlstyle{rm}\Url}\fi

\bibitem[Aso et~al.(2013)Aso, Kan, Shimakawa, and
  Kurata]{Aso2013_scirep_octahedral}
Aso, R., Kan, D., Shimakawa, Y., and Kurata, H., (2013).
\newblock {Atomic level observation of octahedral distortions at the perovskite
  oxide heterointerface}.
\newblock \emph{Sci. Rep.}, {\bfseries 3}, \penalty0 2214.
\newblock \doi{10.1038/srep02214}.

\bibitem[Azough et~al.(2016)Azough, Cernik, Schaffer, Kepaptsoglou, Ramasse,
  Bigatti, Ali, MacLaren, Barthel, Molinari, Baran, Parker, and
  Freer]{Azough_TB_2016}
Azough, F., Cernik, R.~J., Schaffer, B., Kepaptsoglou, D., Ramasse, Q.~M.,
  Bigatti, M., Ali, A., MacLaren, I., Barthel, J., Molinari, M., Baran, J.~D.,
  Parker, S.~C., and Freer, R., (2016).
\newblock {Tungsten Bronze Barium Neodymium Titanate
  (Ba$_{6-3n}$Nd$_{8+2n}$Ti$_{18}$O$_{54}$): An Intrinsic Nanostructured
  Material and Its Defect Distribution}.
\newblock \emph{Inorg. Chem.}, {\bfseries {\bfseries 55}\penalty0 (7)},
  \penalty0 3338--3350.
\newblock \doi{10.1021/acs.inorgchem.5b02594}.

\bibitem[Ballabriga et~al.(2013)Ballabriga, Alozy, Blaj, Campbell, Fiederle,
  Frojdh, Heijne, Llopart, Pichotka, Procz, Tlustos, and
  Wong]{Ballabriga_2013_Medipix3RX}
Ballabriga, R., Alozy, J., Blaj, G., Campbell, M., Fiederle, M., Frojdh, E.,
  Heijne, E. H.~M., Llopart, X., Pichotka, M., Procz, S., Tlustos, L., and
  Wong, W., (feb 2013).
\newblock {The Medipix3RX: a high resolution, zero dead-time pixel detector
  readout chip allowing spectroscopic imaging}.
\newblock \emph{J. Instrum.}, {\bfseries {\bfseries 8}\penalty0 (02)},
  \penalty0 C02016--C02016.
\newblock \doi{10.1088/1748-0221/8/02/c02016}.

\bibitem[Bashir et~al.(2019)Bashir, Millar, Gallacher, Paul, Darbal, Stroud,
  Ballabio, Frigerio, Isella, and MacLaren]{Bashir_JAP_2019_sped_Ge}
Bashir, A., Millar, R.~W., Gallacher, K., Paul, D.~J., Darbal, A.~D., Stroud,
  R., Ballabio, A., Frigerio, J., Isella, G., and MacLaren, I., (2019).
\newblock {Strain analysis of a Ge micro disk using precession electron
  diffraction}.
\newblock \emph{J. Appl. Phys.}, {\bfseries {\bfseries 126}\penalty0 (23)},
  \penalty0 235701.
\newblock \doi{10.1063/1.5113761}.

\bibitem[B\'{e}ch\'{e} et~al.(2009)B\'{e}ch\'{e}, Rouvi\`{e}re, Cl\'{e}ment,
  and Hartmann]{nbed_beche_2009}
B\'{e}ch\'{e}, A., Rouvi\`{e}re, J.~L., Cl\'{e}ment, L., and Hartmann, J.~M.,
  (2009).
\newblock Improved precision in strain measurement using nanobeam electron
  diffraction.
\newblock \emph{Appl. Phys. Lett.}, {\bfseries {\bfseries 95}\penalty0 (12)},
  \penalty0 123114.
\newblock \doi{10.1063/1.3224886}.

\bibitem[B\'{e}ch\'{e} et~al.(2011)B\'{e}ch\'{e}, Rouvi\`{e}re, Barnes, and
  Cooper]{Beche_2011_dfeh}
B\'{e}ch\'{e}, A., Rouvi\`{e}re, J., Barnes, J., and Cooper, D., (2011).
\newblock Dark field electron holography for strain measurement.
\newblock \emph{Ultramicroscopy}, {\bfseries {\bfseries 111}\penalty0 (3)},
  \penalty0 227--238.
\newblock \doi{10.1016/j.ultramic.2010.11.030}.

\bibitem[B\'{e}ch\'{e} et~al.(2013)B\'{e}ch\'{e}, Rouvi\`{e}re, Barnes, and
  Cooper]{Beche2013_um_strain_comparison}
B\'{e}ch\'{e}, A., Rouvi\`{e}re, J.~L., Barnes, J.~P., and Cooper, D., (2013).
\newblock {Strain measurement at the nanoscale: Comparison between convergent
  beam electron diffraction, nano-beam electron diffraction, high resolution
  imaging and dark field electron holography}.
\newblock \emph{Ultramicroscopy}, {\bfseries 131}, \penalty0 10 -- 23.
\newblock \doi{10.1016/j.ultramic.2013.03.014}.

\bibitem[Borisevich et~al.(2010)Borisevich, Ovchinnikov, Chang, Oxley, Yu,
  Seidel, Eliseev, Morozovska, Ramesh, Pennycook, and
  Kalinin]{Borisevich_AtomicShape_2010}
Borisevich, A.~Y., Ovchinnikov, O.~S., Chang, H.~J., Oxley, M.~P., Yu, P.,
  Seidel, J., Eliseev, E.~A., Morozovska, A.~N., Ramesh, R., Pennycook, S.~J.,
  and Kalinin, S.~V., (2010).
\newblock {Mapping Octahedral Tilts and Polarization Across a Domain Wall in
  BiFeO$_3$ from Z-Contrast Scanning Transmission Electron Microscopy Image
  Atomic Column Shape Analysis}.
\newblock \emph{ACS Nano}, {\bfseries {\bfseries 4}\penalty0 (10)}, \penalty0
  6071--6079.
\newblock \doi{10.1021/nn1011539}.

\bibitem[B\"ucker et~al.(2020)B\"ucker, Hogan-Lamarre, Mehrabi, Schulz,
  Bultema, Gevorkov, Brehm, Yefanov, Oberth\"ur, Kassier, and
  Dwayne~Miller]{Bucker_2020_nat_commun_mpx_tile_struct}
B\"ucker, R., Hogan-Lamarre, P., Mehrabi, P., Schulz, E.~C., Bultema, L.~A.,
  Gevorkov, Y., Brehm, W., Yefanov, O., Oberth\"ur, D., Kassier, G.~H., and
  Dwayne~Miller, R.~J., (2020).
\newblock Serial protein crystallography in an electron microscope.
\newblock \emph{Nat. Commun.}, {\bfseries 11}, \penalty0 996.
\newblock \doi{10.1038/s41467-020-14793-0}.

\bibitem[Clabbers et~al.(2017)Clabbers, van Genderen, Wan, Wiegers, Gruene, and
  Abrahams]{Clabbers_2017_ACD_mpx_tile_struct}
Clabbers, M. T.~B., van Genderen, E., Wan, W., Wiegers, E.~L., Gruene, T., and
  Abrahams, J.~P., (2017).
\newblock Protein structure determination by electron diffraction using a
  single three-dimensional nanocrystal.
\newblock \emph{Acta Crystallogr. D}, {\bfseries 73}, \penalty0 738 -- 748.
\newblock \doi{10.1107/S2059798317010348}.

\bibitem[Clausen et~al.(2019)Clausen, Weber, @probonopd, Caron, Nord,
  Müller-Caspary, Ophus, Dunin-Borkowski, Ruzaeva, Chandra, Shin, and van
  Schyndel]{libertem}
Clausen, A., Weber, D., @probonopd, Caron, J., Nord, M., Müller-Caspary, K.,
  Ophus, C., Dunin-Borkowski, R., Ruzaeva, K., Chandra, R., Shin, J., and van
  Schyndel, J., (October 2019).
\newblock Libertem/libertem: 0.2.2.
\newblock \url{https://doi.org/10.5281/zenodo.3489385}.

\bibitem[Collette(2013)]{collette_python_hdf5_2014}
Collette, A., (2013).
\newblock \emph{Python and HDF5}.
\newblock O'Reilly.

\bibitem[Cooper et~al.(2009)Cooper, Barnes, Hartmann, B\'{e}ch\'{e}, and
  Rouvi\`{e}re]{Cooper_2009_apl_dfeh}
Cooper, D., Barnes, J., Hartmann, J., B\'{e}ch\'{e}, A., and Rouvi\`{e}re, J.,
  (2009).
\newblock Dark field electron holography for quantitative strain measurements
  with nanometer-scale spatial resolution.
\newblock \emph{Appl. Phys. Lett.}, {\bfseries {\bfseries 95}\penalty0 (5)},
  \penalty0 053501.
\newblock \doi{10.1063/1.3196549}.

\bibitem[Cooper et~al.(2016)Cooper, Denneulin, Bernier, B\'{e}ch\'{e}, and
  Rouvi\`{e}re]{Cooper_2016_semi_strain_micron}
Cooper, D., Denneulin, T., Bernier, N., B\'{e}ch\'{e}, A., and Rouvi\`{e}re,
  J.-L., (2016).
\newblock Strain mapping of semiconductor specimens with nm-scale resolution in
  a transmission electron microscope.
\newblock \emph{Micron}, {\bfseries 80}, \penalty0 145 -- 165.
\newblock \doi{10.1016/j.micron.2015.09.001}.

\bibitem[{Dask~Development~Team}(2016)]{dask_library}
{Dask~Development~Team}, (2016).
\newblock Dask: Library for dynamic task scheduling.
\newblock \url{http://dask.pydata.org}.

\bibitem[de~la Pe{\~n}a et~al.(2018)de~la Pe{\~n}a, Ostasevicius, Fauske,
  Burdet, Prestat, Jokubauskas, Nord, Sarahan, MacArthur, Johnstone, Taillon,
  Caron, Migunov, Furnival, Eljarrat, Mazzucco, Aarholt, Walls, Slater,
  Winkler, Martineau, Donval, McLeod, Hoglund, Alxneit, Hjorth, Henninen,
  Zagonel, Garmannslund, and 5ht2]{hyperspy_library}
de~la Pe{\~n}a, F., Ostasevicius, T., Fauske, V.~T., Burdet, P., Prestat, E.,
  Jokubauskas, P., Nord, M., Sarahan, M., MacArthur, K.~E., Johnstone, D.~N.,
  Taillon, J., Caron, J., Migunov, V., Furnival, T., Eljarrat, A., Mazzucco,
  S., Aarholt, T., Walls, M., Slater, T., Winkler, F., Martineau, B., Donval,
  G., McLeod, R., Hoglund, E.~R., Alxneit, I., Hjorth, I., Henninen, T.,
  Zagonel, L.~F., Garmannslund, A., and 5ht2, (2018).
\newblock hyperspy/hyperspy: {HyperSpy} 1.3.1.
\newblock \url{https://doi.org/10.5281/zenodo.1221347}.

\bibitem[Donoho and Johnstone(1994)]{Donoho_1994_wavelet_noise_est}
Donoho, D.~L. and Johnstone, I.~M., (1994).
\newblock {Ideal spatial adaptation by wavelet shrinkage}.
\newblock \emph{Biometrika}, {\bfseries 81}, \penalty0 425--455.
\newblock \doi{10.1093/biomet/81.3.425}.

\bibitem[Egerton(2011)]{Egerton_EELS}
Egerton, R.~F., (2011).
\newblock \emph{Electron Energy-Loss Spectroscopy in the Electron Microscope}.
\newblock Springer New York, third edition.

\bibitem[{EMD~authors}(2019)]{emd_format}
{EMD~authors}, (2019).
\newblock {Electron Microscopy Datasets: An HDF5-based interchange file format
  for electron microscopy data and metadata}.
\newblock \url{https://emdatasets.com/format}.
\newblock {Accessed} June 3, 2018.

\bibitem[Emslie(1934)]{Emslie_ElectronDiffraction_1934}
Emslie, A.~G., (1934).
\newblock {Scattering of electrons by stibnite and galena}.
\newblock \emph{{Phys. Rev.}}, {\bfseries {\bfseries 45}\penalty0 (1)},
  \penalty0 43--46.
\newblock \doi{{10.1103/PhysRev.45.43}}.

\bibitem[Findlay et~al.(2010)Findlay, Shibata, Sawada, Okunishi, Kondo, and
  Ikuhara]{FINDLAY_2010_UltraMicros_ABF}
Findlay, S.~D., Shibata, N., Sawada, H., Okunishi, E., Kondo, Y., and Ikuhara,
  Y., (2010).
\newblock Dynamics of annular bright field imaging in scanning transmission
  electron microscopy.
\newblock \emph{Ultramicroscopy}, {\bfseries {\bfseries 110}\penalty0 (7)},
  \penalty0 903 -- 923.
\newblock \doi{10.1016/j.ultramic.2010.04.004}.

\bibitem[{fpd~demos~devs}(2018)]{fpd_demos}
{fpd~demos~devs}, (2018).
\newblock {Notebook examples for the fpd package}.
\newblock \url{https://gitlab.com/fpdpy/fpd-demos}.
\newblock {Accessed} June 3, 2018.

\bibitem[{fpd~devs}(2015)]{fpd}
{fpd~devs}, (2015).
\newblock {FPD: Fast pixelated detector data storage, analysis and
  visualisation}.
\newblock \url{https://gitlab.com/fpdpy/fpd}.
\newblock {Accessed} February 6, 2018.

\bibitem[Friedel(1913)]{Friedel_1913}
Friedel, G., (1913).
\newblock Sur les sym{\'e}tries cristallines que peut r{\'e}v{\'e}ler la
  diffraction des rayons r{\"o}ntgen.
\newblock \emph{C. R. Acad. Sci.}, {\bfseries 157}, \penalty0 1533--1536.

\bibitem[Gammer et~al.(2015)Gammer, Ozdol, Liebscher, and
  Minor]{Gammer_UM_2015_virt_ap}
Gammer, C., Ozdol, V.~B., Liebscher, C.~H., and Minor, A.~M., (2015).
\newblock Diffraction contrast imaging using virtual apertures.
\newblock \emph{Ultramicroscopy}, {\bfseries 155}, \penalty0 1 -- 10.
\newblock \doi{10.1016/j.ultramic.2015.03.015}.

\bibitem[Glazer(1972)]{Glazer_TiltingPerovskites_1972}
Glazer, A.~M., (1972).
\newblock The classification of tilted octahedra in perovskites.
\newblock \emph{Acta Crystallogr. Sec. B}, {\bfseries {\bfseries 28}\penalty0
  (11)}, \penalty0 3384--3392.
\newblock \doi{10.1107/S0567740872007976}.

\bibitem[Gouillart et~al.(2016)Gouillart, Nunez-Iglesias, and van~der
  Walt]{Gouillart_2016_python_scikit}
Gouillart, E., Nunez-Iglesias, J., and van~der Walt, S., (2016).
\newblock Analyzing microtomography data with {Python} and the scikit-image
  library.
\newblock \emph{Adv. Struct. Chem. Imaging}, {\bfseries {\bfseries 2}\penalty0
  (1)}, \penalty0 18.
\newblock \doi{10.1186/s40679-016-0031-0}.

\bibitem[Guzzinati et~al.(2019)Guzzinati, Ghielens, Mahr, B{\'e}ch{\'e},
  Rosenauer, Calders, and Verbeeck]{Guzzinati_2019_bessel_arxiv}
Guzzinati, G., Ghielens, W., Mahr, C., B{\'e}ch{\'e}, A., Rosenauer, A.,
  Calders, T., and Verbeeck, J., (2019).
\newblock {Electron Bessel beam diffraction for precise and accurate nanoscale
  strain mapping}.
\newblock \url{arXiv:1902.06979}.

\bibitem[Hallsteinsen et~al.(2016)Hallsteinsen, Moreau, Grutter, Nord, Vullum,
  Gilbert, Bolstad, Grepstad, Holmestad, Selbach, N'Diaye, Kirby, Arenholz, and
  Tybell]{ingrid_lsmo_lfo_sto_2016}
Hallsteinsen, I., Moreau, M., Grutter, A., Nord, M., Vullum, P.-E., Gilbert,
  D.~A., Bolstad, T., Grepstad, J.~K., Holmestad, R., Selbach, S.~M., N'Diaye,
  A.~T., Kirby, B.~J., Arenholz, E., and Tybell, T., (2016).
\newblock Concurrent magnetic and structural reconstructions at the interface
  of (111)-oriented $\mathrm{La_{0.7}Sr_{0.3}MnO_{3}} \text{/}
  \mathrm{LaFeO_{3}}$.
\newblock \emph{Phys. Rev. B}, {\bfseries 94}, \penalty0 201115.
\newblock \doi{10.1103/PhysRevB.94.201115}.

\bibitem[Hammel and Rose(1995)]{HAMMEL_1995_UltraMicros_ABF}
Hammel, M. and Rose, H., (1995).
\newblock Optimum rotationally symmetric detector configurations for
  phase-contrast imaging in scanning transmission electron microscopy.
\newblock \emph{Ultramicroscopy}, {\bfseries {\bfseries 58}\penalty0 (3)},
  \penalty0 403 -- 415.
\newblock \doi{10.1016/0304-3991(95)00007-N}.

\bibitem[Hart et~al.(2016)Hart, Bassiri, Borisenko, V\'{e}ron, Rauch, Martin,
  Rowan, Fejer, and MacLaren]{Hart_2016_fem}
Hart, M.~J., Bassiri, R., Borisenko, K.~B., V\'{e}ron, M., Rauch, E.~F.,
  Martin, I.~W., Rowan, S., Fejer, M.~M., and MacLaren, I., (2016).
\newblock Medium range structural order in amorphous tantala spatially resolved
  with changes to atomic structure by thermal annealing.
\newblock \emph{J. Non-Cryst. Solids}, {\bfseries 438}, \penalty0 10 -- 17.
\newblock \doi{10.1016/j.jnoncrysol.2016.02.005}.

\bibitem[Hartel et~al.(1996)Hartel, Rose, and
  Dinges]{HARTEL_1996_UltraMicros_haadf}
Hartel, P., Rose, H., and Dinges, C., (1996).
\newblock Conditions and reasons for incoherent imaging in {STEM}.
\newblock \emph{Ultramicroscopy}, {\bfseries {\bfseries 63}\penalty0 (2)},
  \penalty0 93 -- 114.
\newblock \doi{10.1016/0304-3991(96)00020-4}.

\bibitem[Huang et~al.(2010)Huang, Gloter, Chu, Chou, Shu, Liu, Chen, and
  Colliex]{Huang_holz_stem_2010}
Huang, F.-T., Gloter, A., Chu, M.-W., Chou, F.~C., Shu, G.~J., Liu, L.-K.,
  Chen, C.~H., and Colliex, C., (2010).
\newblock Scanning transmission electron microscopy using selective high-order
  {Laue} zones: Three-dimensional atomic ordering in sodium cobaltate.
\newblock \emph{Phys. Rev. Lett.}, {\bfseries 105}, \penalty0 125502.
\newblock \doi{10.1103/PhysRevLett.105.125502}.

\bibitem[Hunter(2007)]{matplotlib}
Hunter, J.~D., (2007).
\newblock Matplotlib: A {2D} graphics environment.
\newblock \emph{Comput. Sci. Eng.}, {\bfseries {\bfseries 9}\penalty0 (3)},
  \penalty0 90--95.
\newblock \doi{10.1109/MCSE.2007.55}.

\bibitem[H{\"{y}}tch et~al.(2008)H{\"{y}}tch, H{\"{u}}e, and
  Snoeck]{Hytch2008_dfeh}
H{\"{y}}tch, F., M.~Houdellier, H{\"{u}}e, F., and Snoeck, E., (2008).
\newblock {Nanoscale holographic interferometry for strain measurements in
  electronic devices}.
\newblock \emph{Nature}, {\bfseries 453}, \penalty0 1086--1089.
\newblock \doi{10.1038/nature07049}.

\bibitem[H{\"{y}}tch et~al.(1998)H{\"{y}}tch, Snoeck, and
  Kilaas]{Hytch1998_gpa}
H{\"{y}}tch, M.~J., Snoeck, E., and Kilaas, R., (1998).
\newblock {Quantitative measurement of displacement and strain fields from HREM
  micrographs}.
\newblock \emph{Ultramicroscopy}, {\bfseries {\bfseries 74}\penalty0 (3)},
  \penalty0 131--146.
\newblock \doi{10.1016/S0304-3991(98)00035-7}.

\bibitem[Jin et~al.(2015)Jin, Nori, Lee, Kumar, Wu, Prater, Kim, and
  Narayan]{Zhenghe_JAP_2015_MgO}
Jin, Z., Nori, S., Lee, Y.-F., Kumar, D., Wu, F., Prater, J.~T., Kim, K.~W.,
  and Narayan, J., (2015).
\newblock Strain induced room temperature ferromagnetism in epitaxial magnesium
  oxide thin films.
\newblock \emph{J. Appl. Phys.}, {\bfseries {\bfseries 118}\penalty0 (16)},
  \penalty0 165309.
\newblock \doi{10.1063/1.4934498}.

\bibitem[Johnstone et~al.(2019)Johnstone, Crout, H{\o}g{\aa}s, Martineau,
  Smeets, Laulainen, Collins, Morzy, Prestat, {\AA}nes, phillipcrout, Doherty,
  Ostasevicius, and Bergh]{pyxem}
Johnstone, D.~N., Crout, P., H{\o}g{\aa}s, S., Martineau, B., Smeets, S.,
  Laulainen, J., Collins, S., Morzy, J., Prestat, E., {\AA}nes, H.,
  phillipcrout, Doherty, T., Ostasevicius, T., and Bergh, T., (September 2019).
\newblock pyxem/pyxem: pyxem 0.9.2.
\newblock \url{https://doi.org/10.5281/zenodo.3407316}.

\bibitem[Jones et~al.(2001)Jones, Oliphant, Peterson, et~al.]{scipy}
Jones, E., Oliphant, T., Peterson, P., et~al., (2001).
\newblock {SciPy}: Open source scientific tools for {Python}.
\newblock \url{http://www.scipy.org}.
\newblock {Accessed} 30/10/2018.

\bibitem[Jones et~al.(2018)Jones, Varambhia, Sawada, and
  Nellist]{Jones_JoM_2018_stem_det_cal}
Jones, L., Varambhia, A., Sawada, H., and Nellist, P.~D., (2018).
\newblock An optical configuration for fastidious {STEM} detector calibration
  and the effect of the objective-lens pre-field.
\newblock \emph{J. Microsc.}, {\bfseries {\bfseries 270}\penalty0 (2)},
  \penalty0 176--187.
\newblock \doi{10.1111/jmi.12672}.

\bibitem[Jones et~al.({1977})Jones, Rackham, and
  Steeds]{HOLZ_Jones_Rackham_Steeds_1977}
Jones, P.~M., Rackham, G.~M., and Steeds, J.~W., ({1977}).
\newblock {Higher-order Laue zone effects in electron-diffraction and their use
  in lattice-parameter determination}.
\newblock \emph{{Proc. R. Soc. London Ser. A}}, {\bfseries {\bfseries
  {354}}\penalty0 (1677)}, \penalty0 197.
\newblock \doi{10.1098/rspa.1977.0064}.

\bibitem[Kim et~al.(2017)Kim, Pennycook, and
  Borisevich]{Kim2017_um_oxygen_octahedral}
Kim, Y.-M., Pennycook, S.~J., and Borisevich, A.~Y., (2017).
\newblock Quantitative comparison of bright field and annular bright field
  imaging modes for characterization of oxygen octahedral tilts.
\newblock \emph{Ultramicroscopy}, {\bfseries 181}, \penalty0 1 -- 7.
\newblock \doi{10.1016/j.ultramic.2017.04.020}.

\bibitem[Klinger(2017)]{Klinger17_crystbox}
Klinger, M., (2017).
\newblock {More features, more tools, more {\it CrysTBox}}.
\newblock \emph{J. Appl. Crystallogr.}, {\bfseries {\bfseries 50}\penalty0
  (4)}, \penalty0 1226--1234.
\newblock \doi{10.1107/S1600576717006793}.

\bibitem[Kluyver et~al.(2016)Kluyver, Ragan-Kelley, P{\'e}rez, Granger,
  Bussonnier, Frederic, Kelley, Hamrick, Grout, Corlay, Ivanov, Avila, Abdalla,
  and Willing]{Kluyver_2016_jupyter}
Kluyver, T., Ragan-Kelley, B., P{\'e}rez, F., Granger, B., Bussonnier, M.,
  Frederic, J., Kelley, K., Hamrick, J., Grout, J., Corlay, S., Ivanov, P.,
  Avila, D., Abdalla, S., and Willing, C., (2016).
\newblock Jupyter notebooks -- a publishing format for reproducible
  computational workflows.
\newblock In Loizides, F. and Schmidt, B., editors, \emph{Positioning and Power
  in Academic Publishing: Players, Agents and Agendas}, pages 87 -- 90. IOS
  Press.

\bibitem[Kolb et~al.(2007)Kolb, Gorelik, Kübel, Otten, and
  Hubert]{Kolb_2007_um_sped_tomo}
Kolb, U., Gorelik, T., Kübel, C., Otten, M., and Hubert, D., (2007).
\newblock {Towards automated diffraction tomography: Part I—Data
  acquisition}.
\newblock \emph{Ultramicroscopy}, {\bfseries 107}, \penalty0 507 -- 513.
\newblock \doi{10.1016/j.ultramic.2006.10.007}.

\bibitem[Krajnak et~al.(2016)Krajnak, McGrouther, M., O'Shea, and
  McVitie]{matus_pixelated_stem_magnetic_2016}
Krajnak, M., McGrouther, D., M., D., O'Shea, V., and McVitie, S., (2016).
\newblock Pixelated detectors and improved efficiency for magnetic imaging in
  {STEM} differential phase contrast.
\newblock \emph{Ultramicroscopy}, {\bfseries 165}, \penalty0 42 -- 50.
\newblock \doi{10.1016/j.ultramic.2016.03.006}.

\bibitem[Krakow and Howland(1976)]{KRAKOW_1976_hollow_cone}
Krakow, W. and Howland, L.~A., (1976).
\newblock A method for producing hollow cone illumination electronically in the
  conventional transmission microscope.
\newblock \emph{Ultramicroscopy}, {\bfseries 2}, \penalty0 53 -- 67.
\newblock \doi{10.1016/S0304-3991(76)90416-2}.

\bibitem[LeBeau et~al.(2009)LeBeau, D'Alfonso, Findlay, Stemmer, and
  Allen]{LeBeau_2009_PRB_BF}
LeBeau, J.~M., D'Alfonso, A.~J., Findlay, S.~D., Stemmer, S., and Allen, L.~J.,
  (2009).
\newblock Quantitative comparisons of contrast in experimental and simulated
  bright-field scanning transmission electron microscopy images.
\newblock \emph{Phys. Rev. B}, {\bfseries 80}, \penalty0 174106.
\newblock \doi{10.1103/PhysRevB.80.174106}.

\bibitem[Lewis et~al.(2016)Lewis, Marrows, and Langridge]{Lewis_2016_FeRh_rev}
Lewis, L.~H., Marrows, C.~H., and Langridge, S., (2016).
\newblock Coupled magnetic, structural, and electronic phase transitions in
  {FeRh}.
\newblock \emph{J. Phys. D Appl. Phys.}, {\bfseries {\bfseries 49}\penalty0
  (32)}, \penalty0 323002.
\newblock \doi{10.1088/0022-3727/49/32/323002}.

\bibitem[MacLaren and Richter(2009)]{MacLarenRichter_2009}
MacLaren, I. and Richter, G., (2009).
\newblock Structure and possible origins of stacking faults in gamma-yttrium
  disilicate.
\newblock \emph{Philos. Mag.}, {\bfseries {\bfseries 89}\penalty0 (2)},
  \penalty0 169--181.
\newblock \doi{10.1080/14786430802562132}.

\bibitem[MacLaren et~al.(2013)MacLaren, Wang, Morris, Craven, Stamps, Schaffer,
  Ramasse, Miao, Kalantari, Sterianou, and
  Reaney]{MacLaren_2015_APLmat_haadf_bf}
MacLaren, I., Wang, L., Morris, O., Craven, A.~J., Stamps, R.~L., Schaffer, B.,
  Ramasse, Q.~M., Miao, S., Kalantari, K., Sterianou, I., and Reaney, I.~M.,
  (2013).
\newblock {Local stabilisation of polar order at charged antiphase boundaries
  in antiferroelectric (Bi$_{0.85}$Nd$_{0.15}$)(Ti$_{0.1}$Fe$_{0.9}$)O$_3$}.
\newblock \emph{APL Mater.}, {\bfseries {\bfseries 1}\penalty0 (2)}, \penalty0
  021102.
\newblock \doi{10.1063/1.4818002}.

\bibitem[MacLaren et~al.(2020)MacLaren, Frutos-Myro, McGrouther, McFadzean,
  Weiss, Cosart, Portillo, Robins, Nicolopoulos, del Busto, and
  Skogeby]{MacLaren_MandM_2020_sped}
MacLaren, I., Frutos-Myro, E., McGrouther, D., McFadzean, S., Weiss, J.~K.,
  Cosart, D., Portillo, J., Robins, A., Nicolopoulos, S., del Busto, E.~N., and
  Skogeby, R., (2020).
\newblock Orientation mapping using scanned precession electron diffraction
  with a direct electron detector.
\newblock \emph{Microsc. Microanal.}, {\bfseries TBA}, \penalty0 TBA.
\newblock \doi{TBA}.

\bibitem[Mahr et~al.(2019)Mahr, M{\"u}ller-Caspary, Ritz, Simson, Grieb,
  Schowalter, Krause, Lackmann, Soltau, Wittstock, and
  Rosenauer]{knut_muller_nbed_2019}
Mahr, C., M{\"u}ller-Caspary, K., Ritz, R., Simson, M., Grieb, T., Schowalter,
  M., Krause, F.~F., Lackmann, A., Soltau, H., Wittstock, A., and Rosenauer,
  A., (2019).
\newblock Influence of distortions of recorded diffraction patterns on strain
  analysis by nano-beam electron diffraction.
\newblock \emph{Ultramicroscopy}, {\bfseries 196}, \penalty0 74 -- 82.
\newblock \doi{10.1016/j.ultramic.2018.09.010}.

\bibitem[Martineau et~al.(2019)Martineau, Johnstone, van Helvoort, Midgley, and
  Eggeman]{Martineau_ASCI_2019_sped_machine}
Martineau, B.~H., Johnstone, D.~N., van Helvoort, A. T.~J., Midgley, P.~A., and
  Eggeman, A.~S., (2019).
\newblock Unsupervised machine learning applied to scanning precession electron
  diffraction data.
\newblock \emph{Adv. Struct. and Chem. Imaging}, {\bfseries {\bfseries
  5}\penalty0 (1)}, \penalty0 3.
\newblock \doi{10.1186/s40679-019-0063-3}.

\bibitem[Mawson et~al.(2020)Mawson, Nakamura, Petersen, Shibata, Sasakif,
  Paganin, Morgan, and Findlay]{Mawson_2020_sped_DPC}
Mawson, T., Nakamura, A., Petersen, T.~C., Shibata, N., Sasakif, H., Paganin,
  D.~M., Morgan, M.~J., and Findlay, S., (2020).
\newblock {Suppressing dynamical diffraction artefacts in differential phase
  contrast scanning transmission electron microscopy of long-range
  electromagnetic fields via precession}.
\newblock \url{arXiv:2002.01595}.

\bibitem[McMullan et~al.(2007)McMullan, Cattermole, Chen, Henderson, Llopart,
  Summerfield, Tlustos, and Faruqi]{McMullan2007_mpx2_electron_imaging}
McMullan, G., Cattermole, D., Chen, S., Henderson, R., Llopart, X.,
  Summerfield, C., Tlustos, L., and Faruqi, A., (2007).
\newblock {Electron imaging with Medipix2 hybrid pixel detector}.
\newblock \emph{Ultramicroscopy}, {\bfseries {\bfseries 107}\penalty0 (4)},
  \penalty0 401 -- 413.
\newblock \doi{10.1016/j.ultramic.2006.10.005}.

\bibitem[McVitie et~al.(2015)McVitie, McGrouther, McFadzean, MacLaren, O'Shea,
  and Benitez]{mcvitie2015_magtem}
McVitie, S., McGrouther, D., McFadzean, S., MacLaren, D.~A., O'Shea, K.~J., and
  Benitez, M.~J., (2015).
\newblock Aberration corrected {Lorentz} scanning transmission electron
  microscopy.
\newblock \emph{Ultramicroscopy}, {\bfseries 152}, \penalty0 57 -- 62.
\newblock \doi{10.1016/j.ultramic.2015.01.003}.

\bibitem[Midgley and Eggeman(2015)]{Midgley_IUCrJ_2015_ped_rev}
Midgley, P.~A. and Eggeman, A.~S., (2015).
\newblock {Precession electron diffraction {--} a topical review}.
\newblock \emph{IUCrJ}, {\bfseries {\bfseries 2}\penalty0 (1)}, \penalty0
  126--136.
\newblock \doi{10.1107/S2052252514022283}.

\bibitem[Mir et~al.(2017)Mir, Clough, MacInnes, Gough, Plackett, Shipsey,
  Sawada, MacLaren, Ballabriga, Maneuski, O'Shea, McGrouther, and
  Kirkland]{MIR_2017_medipix3_characterisation}
Mir, J.~A., Clough, R., MacInnes, R., Gough, C., Plackett, R., Shipsey, I.,
  Sawada, H., MacLaren, I., Ballabriga, R., Maneuski, D., O'Shea, V.,
  McGrouther, D., and Kirkland, A.~I., (2017).
\newblock Characterisation of the {Medipix3} detector for 60 and {80keV}
  electrons.
\newblock \emph{Ultramicroscopy}, {\bfseries 182}, \penalty0 44 -- 53.
\newblock \doi{10.1016/j.ultramic.2017.06.010}.

\bibitem[Mugnaioli et~al.(2009)Mugnaioli, Gorelik, and
  Kolb]{Mugnaioli2009_um_sped_tomo}
Mugnaioli, E., Gorelik, T., and Kolb, U., (2009).
\newblock {“Ab initio” structure solution from electron diffraction data
  obtained by a combination of automated diffraction tomography and precession
  technique}.
\newblock \emph{Ultramicroscopy}, {\bfseries 109}, \penalty0 758 -- 765.
\newblock \doi{10.1016/j.ultramic.2009.01.011}.

\bibitem[Naden et~al.(2018)Naden, O'Shea, and MacLaren]{Naden2018_moire_stem}
Naden, A.~B., O'Shea, K.~J., and MacLaren, D.~A., (apr 2018).
\newblock {Evaluation of crystallographic strain, rotation and defects in
  functional oxides by the moir{\'{e}} effect in scanning transmission electron
  microscopy}.
\newblock \emph{Nanotechnology}, {\bfseries {\bfseries 29}\penalty0 (16)},
  \penalty0 165704.
\newblock ISSN 0957-4484.
\newblock \doi{10.1088/1361-6528/aaae50}.

\bibitem[Nord et~al.(2017)Nord, Vullum, MacLaren, Tybell, and
  Holmestad]{Nord2017_atomap}
Nord, M., Vullum, P.~E., MacLaren, I., Tybell, T., and Holmestad, R., (2017).
\newblock Atomap: a new software tool for the automated analysis of atomic
  resolution images using two-dimensional {G}aussian fitting.
\newblock \emph{Adv. Struct. Chem. Imaging}, {\bfseries {\bfseries 3}\penalty0
  (1)}, \penalty0 9.
\newblock \doi{10.1186/s40679-017-0042-5}.

\bibitem[Nord et~al.(2019{\natexlab{a}})Nord, Ross, McGrouther, Barthel,
  Moreau, Hallsteinsen, Tybell, and MacLaren]{Nord_2018_holz}
Nord, M., Ross, A., McGrouther, D., Barthel, J., Moreau, M., Hallsteinsen, I.,
  Tybell, T., and MacLaren, I., (2019{\natexlab{a}}).
\newblock Three-dimensional subnanoscale imaging of unit cell doubling due to
  octahedral tilting and cation modulation in strained perovskite thin films.
\newblock \emph{Phys. Rev. Mater.}, {\bfseries 3}, \penalty0 063605.
\newblock \doi{10.1103/PhysRevMaterials.3.063605}.

\bibitem[Nord et~al.(2019{\natexlab{b}})Nord, Webster, Paton, McVitie,
  McGrouther, MacLaren, and Paterson]{fpd_part1_arxiv}
Nord, M., Webster, R. W.~H., Paton, K.~A., McVitie, S., McGrouther, D.,
  MacLaren, I., and Paterson, G.~W., (2019{\natexlab{b}}).
\newblock {Fast Pixelated Detectors in Scanning Transmission Electron
  Microscopy. Part I: Data Acquisition, Live Processing and Storage}.
\newblock \url{arXiv:1911.11560}.

\bibitem[Oliphant(2006)]{numpy}
Oliphant, T.~E., (2006).
\newblock \emph{A guide to {NumPy}}.
\newblock USA: Trelgol Publishing.

\bibitem[Oliphant(2007)]{oliphant_07_python}
Oliphant, T.~E., (2007).
\newblock Python for scientific computing.
\newblock \emph{Comput. Sci. Eng.}, {\bfseries {\bfseries 9}\penalty0 (3)},
  \penalty0 10--20.
\newblock \doi{10.1109/MCSE.2007.58}.

\bibitem[Ophus(2019)]{ophus_mm_2019_4dstem}
Ophus, C., (2019).
\newblock {Four-Dimensional Scanning Transmission Electron Microscopy
  (4D-STEM): From Scanning Nanodiffraction to Ptychography and Beyond}.
\newblock \emph{Microsc. Microanal.}, {\bfseries {\bfseries 25}\penalty0 (3)},
  \penalty0 563–582.
\newblock \doi{10.1017/S1431927619000497}.

\bibitem[Paterson et~al.(2020)Paterson, Webster, Ross, Paton, Macgregor,
  McGrouther, MacLaren, and Nord]{paper_dataset_p2}
Paterson, G.~W., Webster, R. W.~H., Ross, A., Paton, K.~A., Macgregor, T.~A.,
  McGrouther, D., MacLaren, I., and Nord, M., (2020).
\newblock {Dataset}.
\newblock \url{https://doi.org/10.5281/zenodo.3903517}.

\bibitem[Paterson et~al.(2019)Paterson, Macauley, Li, Mac\^edo, Ferguson,
  Morley, Rosamond, Linfield, Marrows, Stamps, and
  McVitie]{Paterson_2019_PRB_asi}
Paterson, G.~W., Macauley, G.~M., Li, Y., Mac\^edo, R., Ferguson, C., Morley,
  S.~A., Rosamond, M.~C., Linfield, E.~H., Marrows, C.~H., Stamps, R.~L., and
  McVitie, S., (2019).
\newblock Heisenberg pseudo-exchange and emergent anisotropies in field-driven
  pinwheel artificial spin ice.
\newblock \emph{Phys. Rev. B}, {\bfseries 100}, \penalty0 174410.
\newblock \doi{10.1103/PhysRevB.100.174410}.

\bibitem[Pekin et~al.(2017)Pekin, Gammer, Ciston, Minor, and
  Ophus]{Pekin_UM_2017_disc_accuracy}
Pekin, T.~C., Gammer, C., Ciston, J., Minor, A.~M., and Ophus, C., (2017).
\newblock Optimizing disk registration algorithms for nanobeam electron
  diffraction strain mapping.
\newblock \emph{Ultramicroscopy}, {\bfseries 176}, \penalty0 170 -- 176.
\newblock \doi{10.1016/j.ultramic.2016.12.021}.

\bibitem[Peng and Gj{\o}nnes(1989)]{Peng_ACA_1989_holz}
Peng, L.-M. and Gj{\o}nnes, J.~K., (1989).
\newblock {Bloch-wave channeling and HOLZ effects in high-energy electron
  diffraction}.
\newblock \emph{Acta Crystallogr. A}, {\bfseries {\bfseries 45}\penalty0 (10)},
  \penalty0 699--703.
\newblock \doi{10.1107/S0108767389005982}.

\bibitem[Pennycook and Jesson(1991)]{Pennycook1991_um}
Pennycook, S. and Jesson, D., (1991).
\newblock High-resolution {Z}-contrast imaging of crystals.
\newblock \emph{Ultramicroscopy}, {\bfseries {\bfseries 37}\penalty0 (1)},
  \penalty0 14 -- 38.
\newblock \doi{10.1016/0304-3991(91)90004-P}.

\bibitem[{pixStem~devs}(2015)]{pixstem}
{pixStem~devs}, (2015).
\newblock {pixStem: Analysis of pixelated {STEM} data}.
\newblock \url{https://gitlab.com/pixstem/pixstem}.
\newblock {Accessed} October 3, 2018.

\bibitem[Plackett et~al.(2013)Plackett, Horswell, Gimenez, Marchal, Omar, and
  Tartoni]{Plackett_2013_merlin}
Plackett, R., Horswell, I., Gimenez, E.~N., Marchal, J., Omar, D., and Tartoni,
  N., (2013).
\newblock Merlin: a fast versatile readout system for {Medipix3}.
\newblock \emph{J. Instrum.}, {\bfseries {\bfseries 8}\penalty0 (01)},
  \penalty0 C01038.
\newblock \doi{10.1088/1748-0221/8/01/C01038}.

\bibitem[Ponchut et~al.(2015)Ponchut, Collet, Herv{\'{e}}, Caer, Cerrai, Siron,
  Dabin, and Ribois]{Ponchut_2015_JoI_tsv_smartpix}
Ponchut, C., Collet, E., Herv{\'{e}}, C., Caer, T.~L., Cerrai, J., Siron, L.,
  Dabin, Y., and Ribois, J.~F., (2015).
\newblock {SMARTPIX, a photon-counting pixel detector for synchrotron
  applications based on Medipix3RX readout chip and active edge pixel sensors}.
\newblock \emph{J. Instrum.}, {\bfseries {\bfseries 10}\penalty0 (01)},
  \penalty0 C01019--C01019.
\newblock \doi{10.1088/1748-0221/10/01/c01019}.

\bibitem[Rauch and Veron(2005)]{Rauch_2005_virt_det}
Rauch, E.~F. and Veron, M., (2005).
\newblock Coupled microstructural observations and local texture measurements
  with an automated crystallographic orientation mapping tool attached to a
  tem.
\newblock \emph{Materialwiss Werkst.}, {\bfseries {\bfseries 36}\penalty0
  (10)}, \penalty0 552--556.
\newblock \doi{10.1002/mawe.200500923}.

\bibitem[Rauch and V\'eron(2014)]{VDF_Rauch_Veron_2014}
Rauch, E.~F. and V\'eron, M., (2014).
\newblock Virtual dark-field images reconstructed from electron diffraction
  patterns.
\newblock \emph{Eur. Phys. J. Appl. Phys.}, {\bfseries {\bfseries 66}\penalty0
  (1)}, \penalty0 10701.
\newblock \doi{10.1051/epjap/2014130556}.

\bibitem[Rauch et~al.(2010)Rauch, Portillo, Nicolopoulos, Bultreys, Rouvimov,
  and Moeck]{Rauch_SPED_2010}
Rauch, E.~F., Portillo, J., Nicolopoulos, S., Bultreys, D., Rouvimov, S., and
  Moeck, P., (2010).
\newblock Automated nanocrystal orientation and phase mapping in the
  transmission electron microscope on the basis of precession electron
  diffraction.
\newblock \emph{Z. Kristallogr.}, {\bfseries 225}, \penalty0 103--109.
\newblock \doi{10.1524/zkri.2010.1205}.

\bibitem[Rouvi{\`e}re and Sarigiannidou(2005)]{Rouviere_UM_2005_gpa}
Rouvi{\`e}re, J.~L. and Sarigiannidou, E., (2005).
\newblock Theoretical discussions on the geometrical phase analysis.
\newblock \emph{Ultramicroscopy}, {\bfseries {\bfseries 106}\penalty0 (1)},
  \penalty0 1 -- 17.
\newblock \doi{10.1016/j.ultramic.2005.06.001}.

\bibitem[Rouvi{\`e}re et~al.(2013)Rouvi{\`e}re, B{\'e}ch{\'e}, Martin,
  Denneulin, and Cooper]{Rouviere_APL_2013_SPED}
Rouvi{\`e}re, J.-L., B{\'e}ch{\'e}, A., Martin, Y., Denneulin, T., and Cooper,
  D., (2013).
\newblock Improved strain precision with high spatial resolution using nanobeam
  precession electron diffraction.
\newblock \emph{Appl. Phys. Lett.}, {\bfseries {\bfseries 103}\penalty0 (24)},
  \penalty0 241913.
\newblock \doi{10.1063/1.4829154}.

\bibitem[{S. S. P. Parkin} et~al.(2004){S. S. P. Parkin}, {C. Kaiser}, {A.
  Panchula}, {P. M. Rice}, {B. Hughes}, {M. Samant}, and {S.-H.
  Yang}]{Parkin_Nature_2004_MgO}
{S. S. P. Parkin}, {C. Kaiser}, {A. Panchula}, {P. M. Rice}, {B. Hughes}, {M.
  Samant}, and {S.-H. Yang}, (2004).
\newblock {Giant tunnelling magnetoresistance at room temperature with MgO
  (100) tunnel barriers}.
\newblock \emph{Nat. Mater.}, {\bfseries {\bfseries 3}\penalty0 (12)},
  \penalty0 862–867.
\newblock \doi{10.1038/nmat1256}.

\bibitem[Savitzky et~al.(2020)Savitzky, Hughes, Zeltmann, Brown, Zhao, Pelz,
  Barnard, Donohue, DaCosta, Pekin, Kennedy, Janish, Schneider, Herring, Gopal,
  Anapolsky, Ercius, Scott, Ciston, Minor, and
  Ophus]{Savitzky_2020_py4dstem_arxiv}
Savitzky, B.~H., Hughes, L.~A., Zeltmann, S.~E., Brown, H.~G., Zhao, S., Pelz,
  P.~M., Barnard, E.~S., Donohue, J., DaCosta, L.~R., Pekin, T.~C., Kennedy,
  E., Janish, M.~T., Schneider, M.~M., Herring, P., Gopal, C., Anapolsky, A.,
  Ercius, P., Scott, M., Ciston, J., Minor, A.~M., and Ophus, C., (2020).
\newblock {py4DSTEM: a software package for multimodal analysis of
  four-dimensional scanning transmission electron microscopy datasets}.
\newblock \url{arXiv:2003.09523}.

\bibitem[Savitzky et~al.(2019)Savitzky, Zeltmann, Barnard, lerandc, Brown,
  Henderson, and Ginsburg]{py4dstem}
Savitzky, B.~H., Zeltmann, S., Barnard, E., lerandc, Brown, H.~G., Henderson,
  M., and Ginsburg, D., (July 2019).
\newblock {py4dstem/py4DSTEM: DOI release}.
\newblock \url{https://doi.org/10.5281/zenodo.3333960}.

\bibitem[Schaffer et~al.(2004)Schaffer, Grogger, and
  Kothleitner]{Schaffer_UM_2004_alignment}
Schaffer, B., Grogger, W., and Kothleitner, G., (2004).
\newblock Automated spatial drift correction for {EFTEM} image series.
\newblock \emph{Ultramicroscopy}, {\bfseries {\bfseries 102}\penalty0 (1)},
  \penalty0 27 -- 36.
\newblock \doi{10.1016/j.ultramic.2004.08.003}.

\bibitem[Schaffer et~al.(2012)Schaffer, Schaffer, and
  Ramasse]{Schaffer2012_um_fib_sample_prep}
Schaffer, M., Schaffer, B., and Ramasse, Q., (2012).
\newblock {Sample preparation for atomic-resolution STEM at low voltages by
  FIB}.
\newblock \emph{Ultramicroscopy}, {\bfseries 114}, \penalty0 62--71.
\newblock \doi{10.1016/j.ultramic.2012.01.005}.

\bibitem[Shibata et~al.(2010)Shibata, Kohno, Findlay, Sawada, Kondo, and
  Ikuhara]{Shibata2010_detector}
Shibata, N., Kohno, Y., Findlay, S.~D., Sawada, H., Kondo, Y., and Ikuhara, Y.,
  (04 2010).
\newblock {New area detector for atomic-resolution scanning transmission
  electron microscopy}.
\newblock \emph{Microscopy}, {\bfseries {\bfseries 59}\penalty0 (6)}, \penalty0
  473--479.
\newblock \doi{10.1093/jmicro/dfq014}.

\bibitem[Somnath et~al.(2019)Somnath, Smith, Laanait, Vasudevan, Ievlev,
  Belianinov, Lupini, Shankar, Kalinin, and Jesse]{pycroscopy}
Somnath, S., Smith, C.~R., Laanait, N., Vasudevan, R.~K., Ievlev, A.,
  Belianinov, A., Lupini, A.~R., Shankar, M., Kalinin, S.~V., and Jesse, S.,
  (2019).
\newblock {USID and Pycroscopy -- Open frameworks for storing and analyzing
  spectroscopic and imaging data}.
\newblock \url{arXiv:1903.09515}.

\bibitem[Spence and Koch(2001)]{SpenceKoch_HOLZ_2001}
Spence, J. and Koch, C., (2001).
\newblock On the measurement of dislocation core periods by nanodiffraction.
\newblock \emph{{Philos. Mag. B}}, {\bfseries {\bfseries 81}\penalty0 (11)},
  \penalty0 1701--1711.
\newblock \doi{10.1080/13642810108223113}.

\bibitem[Spence et~al.(1989)Spence, Zuo, and Lynch]{SpenceZuo_HOLZ_1989}
Spence, J., Zuo, J., and Lynch, J., (1989).
\newblock {On the HOLZ contribution to STEM lattice images formed using
  high-angle dark-field detectors}.
\newblock \emph{Ultramicroscopy}, {\bfseries {\bfseries 31}\penalty0 (2)},
  \penalty0 233 -- 239.
\newblock \doi{10.1016/0304-3991(89)90218-0}.

\bibitem[Su and Zhu(2010)]{Su2010_um_SMF}
Su, D. and Zhu, Y., (2010).
\newblock Scanning moir{\'{e}} fringe imaging by scanning transmission electron
  microscopy.
\newblock \emph{Ultramicroscopy}, {\bfseries {\bfseries 110}\penalty0 (3)},
  \penalty0 229 -- 233.
\newblock \doi{10.1016/j.ultramic.2009.11.015}.

\bibitem[Tate et~al.(2016)Tate, Purohit, Chamberlain, Nguyen, Hovden, Chang,
  Deb, Turgut, Heron, Schlom, Ralph, Fuchs, Shanks, Philipp, Muller, and
  Gruner]{tate_2016_mandm_empad}
Tate, M.~W., Purohit, P., Chamberlain, D., Nguyen, K.~X., Hovden, R., Chang,
  C.~S., Deb, P., Turgut, E., Heron, J.~T., Schlom, D.~G., Ralph, D.~C., Fuchs,
  G.~D., Shanks, K.~S., Philipp, H.~T., Muller, D.~A., and Gruner, S.~M.,
  (2016).
\newblock High dynamic range pixel array detector for scanning transmission
  electron microscopy.
\newblock \emph{Microsc. Microanal.}, {\bfseries {\bfseries 22}\penalty0 (1)},
  \penalty0 237–249.
\newblock \doi{10.1017/S1431927615015664}.

\bibitem[Temple et~al.(2018{\natexlab{a}})Temple, Almeida, Massey, Fallon,
  Lamb, Morley, Maccherozzi, Dhesi, McGrouther, McVitie, Moore, and
  Marrows]{Temple_PRM_2018_FeRh}
Temple, R.~C., Almeida, T.~P., Massey, J.~R., Fallon, K., Lamb, R., Morley,
  S.~A., Maccherozzi, F., Dhesi, S.~S., McGrouther, D., McVitie, S., Moore,
  T.~A., and Marrows, C.~H., (2018{\natexlab{a}}).
\newblock Antiferromagnetic-ferromagnetic phase domain development in
  nanopatterned {FeRh} islands.
\newblock \emph{Phys. Rev. Mater.}, {\bfseries 2}, \penalty0 104406.
\newblock \doi{10.1103/PhysRevMaterials.2.104406}.

\bibitem[Temple et~al.(2018{\natexlab{b}})Temple, Almeida, Massey, Fallon,
  Lamb, Morley, Maccherozzi, Dhesi, McGrouther, McVitie, Moore, and
  Marrows]{Temple_PRM_2018_FeRh_data}
Temple, R.~C., Almeida, T.~P., Massey, J.~R., Fallon, K., Lamb, R., Morley,
  S.~A., Maccherozzi, F., Dhesi, S.~S., McGrouther, D., McVitie, S., Moore,
  T.~A., and Marrows, C.~H., (2018{\natexlab{b}}).
\newblock {4D STEM data set supporting "Antiferromagnetic-ferromagnetic phase
  domain development in nanopatterned FeRh islands"}.
\newblock {University of Glasgow, Enlighten: Research Data}.

\bibitem[{The HDF Group}(1997-2018)]{hdf5_file_format}
{The HDF Group}, (1997-2018).
\newblock {Hierarchical Data Format, version 5}.
\newblock \url{http://www.hdfgroup.org/HDF5}.

\bibitem[Thong et~al.(2001)Thong, Sim, and Phang]{Thong_2001_1imageSNR}
Thong, J. T.~L., Sim, K.~S., and Phang, J. C.~H., (2001).
\newblock Single-image signal-to-noise ratio estimation.
\newblock \emph{Scanning}, {\bfseries {\bfseries 23}\penalty0 (5)}, \penalty0
  328--336.
\newblock \doi{10.1002/sca.4950230506}.

\bibitem[Tick and Campbell(2011)]{Tick_2011_JoI_tsv}
Tick, T. and Campbell, M., (2011).
\newblock {TSV processing of Medipix3 wafers by CEA-LETI: a progress report}.
\newblock \emph{J. Instrum.}, {\bfseries {\bfseries 6}\penalty0 (11)},
  \penalty0 C11018--C11018.
\newblock \doi{10.1088/1748-0221/6/11/c11018}.

\bibitem[Tsai et~al.(2016)Tsai, Chang, Lobato, Van~Dyck, and
  Chen]{Tsai2016_hollow_cone}
Tsai, C.-Y., Chang, Y.-C., Lobato, I., Van~Dyck, D., and Chen, F.-R., (2016).
\newblock Hollow cone electron imaging for single particle {3D} reconstruction
  of proteins.
\newblock \emph{Sci. Rep.}, {\bfseries 6}, \penalty0 27701.
\newblock \doi{10.1038/srep27701}.

\bibitem[Van~Aert et~al.(2011)Van~Aert, Batenburg, Rossell, Erni, and
  Van~Tendeloo]{Van_Aert_DiscreteTomo_2011}
Van~Aert, S., Batenburg, K.~J., Rossell, M.~D., Erni, R., and Van~Tendeloo, G.,
  (2011).
\newblock Three-dimensional atomic imaging of crystalline nanoparticles.
\newblock \emph{Nature}, {\bfseries {\bfseries 470}\penalty0 (7334)}, \penalty0
  374--377.
\newblock \doi{10.1038/nature09741}.

\bibitem[van~der Walt et~al.(2014)van~der Walt, {S}ch\"onberger,
  {Nunez-Iglesias}, {B}oulogne, {W}arner, {Y}ager, {G}ouillart, {Y}u, and the
  scikit-image contributors]{scikit_image}
van~der Walt, S., {S}ch\"onberger, J.~L., {Nunez-Iglesias}, J., {B}oulogne, F.,
  {W}arner, J.~D., {Y}ager, N., {G}ouillart, E., {Y}u, T., and the scikit-image
  contributors, (2014).
\newblock scikit-image: image processing in {P}ython.
\newblock \emph{PeerJ}, {\bfseries 2}, \penalty0 e453.
\newblock \doi{10.7717/peerj.453}.

\bibitem[Vincent and Midgley(1994)]{Vincent_1994_UM_sped}
Vincent, R. and Midgley, P., (1994).
\newblock Double conical beam-rocking system for measurement of integrated
  electron diffraction intensities.
\newblock \emph{Ultramicroscopy}, {\bfseries {\bfseries 53}\penalty0 (3)},
  \penalty0 271 -- 282.
\newblock \doi{10.1016/0304-3991(94)90039-6}.

\bibitem[Voyles and Muller(2002)]{Voyles_2002_UM_fem}
Voyles, P.~M. and Muller, D.~A., (2002).
\newblock {Fluctuation microscopy in the STEM}.
\newblock \emph{Ultramicroscopy}, {\bfseries {\bfseries 93}\penalty0 (2)},
  \penalty0 147 -- 159.
\newblock \doi{10.1016/S0304-3991(02)00155-9}.

\bibitem[Wang et~al.(2016)Wang, Salzberger, Sigle, Suyolcu, and van
  Aken]{Wang2016_um_oxygen_octahedra_picker}
Wang, Y., Salzberger, U., Sigle, W., Suyolcu, Y.~E., and van Aken, P.~A.,
  (2016).
\newblock {Oxygen octahedra picker: A software tool to extract quantitative
  information from STEM images}.
\newblock \emph{Ultramicroscopy}, {\bfseries 168}, \penalty0 46 -- 52.
\newblock \doi{10.1016/j.ultramic.2016.06.001}.

\bibitem[Yakovlev and Libera(2008)]{Yakovlev_Micron_2008_dose_EELS}
Yakovlev, S. and Libera, M., (2008).
\newblock {Dose-limited spectroscopic imaging of soft materials by low-loss
  EELS in the scanning transmission electron microscope}.
\newblock \emph{Micron}, {\bfseries {\bfseries 39}\penalty0 (6)}, \penalty0 734
  -- 740.
\newblock \doi{10.1016/j.micron.2007.10.019}.

\bibitem[Yang et~al.(2015)Yang, Jones, Ryll, Simson, Soltau, Kondo, Sagawa,
  Banba, MacLaren, and Nellist]{Yang_2015_phase_contrast}
Yang, H., Jones, L., Ryll, H., Simson, M., Soltau, H., Kondo, Y., Sagawa, R.,
  Banba, H., MacLaren, I., and Nellist, P.~D., (2015).
\newblock {4D STEM: High efficiency phase contrast imaging using a fast
  pixelated detector}.
\newblock \emph{J. Phys. Conf. Ser.}, {\bfseries 644}, \penalty0 012032.
\newblock \doi{10.1088/1742-6596/644/1/012032}.

\bibitem[Yang et~al.(2006)Yang, Sha, Wang, Wang, and Yang]{Yang_MSEC_2006_MgO}
Yang, Q., Sha, J., Wang, L., Wang, J., and Yang, D., (2006).
\newblock {MgO} nanostructures synthesized by thermal evaporation.
\newblock \emph{Mater. Sci. Eng. C}, {\bfseries {\bfseries 26}\penalty0 (5)},
  \penalty0 1097 -- 1101.
\newblock \doi{10.1016/j.msec.2005.09.082}.

\bibitem[Zeltmann et~al.(2020)Zeltmann, Müller, Bustillo, Savitzky, Hughes,
  Minor, and Ophus]{Zeltmann_UM_2020_patterned_aps}
Zeltmann, S.~E., Müller, A., Bustillo, K.~C., Savitzky, B., Hughes, L., Minor,
  A.~M., and Ophus, C., (2020).
\newblock {Patterned probes for high precision 4D-STEM bragg measurements}.
\newblock \emph{Ultramicroscopy}, {\bfseries 209}, \penalty0 112890.
\newblock \doi{10.1016/j.ultramic.2019.112890}.

\bibitem[Zhu and Park(2006)]{ZHU_Materials_today_2006_MTJ}
Zhu, J.-G. and Park, C., (2006).
\newblock Magnetic tunnel junctions.
\newblock \emph{{Mater. Today}}, {\bfseries {\bfseries 9}\penalty0 (11)},
  \penalty0 36 -- 45.
\newblock \doi{10.1016/S1369-7021(06)71693-5}.

\end{thebibliography}

\onecolumngrid

\end{document}